\documentclass[draftclsnofoot,final]{IEEEtran}

\usepackage{cite}
\usepackage{hyperref}
\usepackage{url}
\usepackage{amssymb,amsmath,amsthm}
\usepackage{array}
\usepackage[small]{caption}
\usepackage{subcaption}
\usepackage{multirow}
\usepackage{psfrag}
\usepackage{cleveref}
\usepackage{color}

\ifCLASSINFOpdf
\else
\usepackage[dvips]{graphicx}
\fi

\makeatletter
\def\th@plain{%
  \thm@notefont{}
  \itshape 
}
\def\th@definition{%
  \thm@notefont{}
  \normalfont 
}
\makeatother

\usepackage[percent]{overpic}
\usepackage{xspace}
\newtheorem{theorem}{Theorem}
\newtheorem{lemma}{Lemma}
\newtheorem{fact}{Fact}
\newtheorem{proposition}{Proposition}
\newtheorem{corollary}{Corollary}
\theoremstyle{definition}
\newtheorem{definition}{Definition}
\theoremstyle{definition}
\newtheorem{remark}{Remark}
\DeclareMathOperator*{\argmax}{arg\,max}

\DeclareMathOperator*{\Ei}{Ei}
\DeclareMathOperator*{\Q}{Q}
\newcommand{\SIR}{\textnormal{SIR}\xspace}
\newcommand{\SNR}{\textnormal{SNR}\xspace}
\newcommand{\SINR}{\textnormal{SINR}\xspace}
\newcommand{\Pc}{\mathsf{P}}
\renewcommand{\P}{\mathbb{P}}
\newcommand{\R}{\mathbb{R}}
\newcommand{\E}{\mathbb{E}}
\newcommand{\ie}{\emph{i.e.,}\@\xspace}
\newcommand{\eg}{\emph{e.g.,}\@\xspace}

\newcommand\numberthis{\addtocounter{equation}{1}\tag{\theequation}}

\renewcommand{\d}{\textnormal{d}}

\def\peq#1{\stackrel{\text{\scriptsize(#1)}}{=}}
\def\pgeq#1{\stackrel{\text{\scriptsize(#1)}}{\geq}}
\def\pleq#1{\stackrel{\text{\scriptsize(#1)}}{\leq}}

\graphicspath{{./figures/}} 

\begin{document}
%
\title{Downlink Cellular Network Analysis with Multi-slope Path Loss Models}

\author{Xinchen~Zhang and Jeffrey G. Andrews\vspace{-0.8cm}
\thanks{Manuscript date \today. 
 
Xinchen Zhang ({\tt xzhang7@alumni.nd.edu}) is now with Qualcomm Inc., San Diego, CA, USA. This work is completed when he was with the University of Texas at Austin, Austin, Texas, USA.

Jeffrey G. Andrews ({\tt jandrews@ece.utexas.edu}) is the corresponding author of this paper and is with the Wireless Networking and Communications Group (WNCG), the Univeristy of Texas at Austin, Austin, Texas, USA.

This work was supported by the Simons Foundation and the National Science Foundation CIF-1016649.}}

\markboth{}{Updated \today}%

\maketitle

\begin{abstract}

Existing cellular network analyses, and even simulations, typically use the standard path loss model where received power decays like $\|x\|^{-\alpha}$ over a distance $\|x\|$.  This standard path loss model is quite idealized, and in most scenarios the path loss exponent $\alpha$ is itself a function of $\|x\|$, typically an increasing one.  Enforcing a single path loss exponent can lead to orders of magnitude differences in average received and interference powers versus the true values.  In this paper we study \emph{multi-slope} path loss models, where different distance ranges are subject to different path loss exponents.   We focus on the dual-slope path loss function, which is a piece-wise power law and continuous and accurately approximates many practical scenarios.  We derive the distributions of SIR, SNR, and finally SINR before finding the potential throughput scaling, which provides insight on the observed cell-splitting rate gain.   The exact mathematical results show that the SIR monotonically decreases with network density, while the converse is true for SNR, and thus the network coverage probability in terms of SINR is maximized at some finite density.   With ultra-densification (network density goes to infinity), there exists a \emph{phase transition} in the near-field path loss exponent $\alpha_0$: if $\alpha_0 >1$ unbounded potential throughput can be achieved asymptotically; if $\alpha_0 <1$, ultra-densification leads in the extreme case to zero throughput.

%

\end{abstract}

\section{Introduction\label{sec:Intro}}


%

A fundamental property of wireless signal transmissions is that their power rapidly decays over distance.  In particular, in free space we know from the Friis equation that over a distance $\| x \|$ the received signal power $P_r$ is given in terms of the transmit power $P_t$ as
\begin{equation}
P_r = P_t G_t G_r  \left( \frac{\lambda_c}{4 \pi \| x \|} \right)^2,
\end{equation}
for a wavelength $\lambda_c$ and antenna transmit and receive gains $G_t$ and $G_r$.   In terrestrial environments, propagation is much more complex to characterize due to ground reflections, scattering, blocking/shadowing, and other physical features.  Since the number of possible realizations of propagation environments is infinite, simplified models stemming from the Friis equation are typically adopted that have at least some measure of empirical support.  A nearly universal characteristic of such models is that the distance dependence is generalized to $\| x \|^{-\alpha}$, with the path loss exponent $\alpha$ being a parameter that can be roughly fit to the environment.   In decibel terms, this gives a form
\begin{equation}
\label{eq:PLintro}
P_r  = P_t + K_0 - 10 \alpha \log_{10} \| x \|,
\end{equation}
where $K_0$ is a catch-all constant that gives the path loss (in dB) at a distance $\| x \| = 1$.  For example, in the Friis equation, $K_0 = G_t G_r (\lambda_c/ 4\pi)^2$.  Hence the slope of the path loss (in dB) is constant and is determined only by $\alpha$ in such a model, which we will term a \emph{standard path loss model}.

\subsection{The Case for Multi-Slope Path Loss Models}

Although the standard path loss model has a great deal of history, and is the basis for most existing cellular network theory, analysis, simulation and design,
it is also known to lead to unrealistic results in some special cases \cite{net:Inaltekin09jsac,net:Dousse04}.
Besides, the standard path loss model does not accurately capture the dependence of the path loss exponent $\alpha$ on the link distance in many important situations.   We now enumerate a few of these as examples, along with the possible consequences to future cellular network optimization and design.  

\textbf{The two-ray model.}  Even a very simple two-ray model with one direct path and one ground-reflected path results in a pronounced dual slope path loss behavior \cite{GoldsmithBook,SchillingMilsteinPickholtzBrunoKanterakisKullbackErcegBiedermanFishmanSalerno1991,FeuersteinBlackardRappaportEtAl1994}.  In particular, with transmit and receive antenna heights of $h_t$ and $h_r$, below a critical distance $R_c \approx 4 h_t h_r / \lambda_c$ the path loss exponent is $\alpha = 2$, while above this distance it changes to $\alpha = 4$.  For a plausible values $h_t = 10$~m, $h_r = 2$~m, and a carrier frequency $f_c = c/\lambda_c = 1$ GHz, we have $R_c=267$~m as a rough threshold.  It is worth emphasizing that there is a massive difference between $\| x \|^{-2}$ and  $\| x \|^{-4}$ for most reasonable values of  $\| x \|$, and that splitting the difference by using $\alpha \approx 3$ results in large errors in both regimes.

\textbf{Dense or clustered networks.}  Wireless networks are rapidly increasing in density, and in doing so are becoming ever more irregular \cite{And13}.  This causes increasing variations in the link distances and the number of appreciable interferers, and makes a ``one size fits all'' path loss model ever more dubious.  For example, a cellular user equipment (UE) might connect with its closest macrocell (or microcell) that is beyond the critical distance, while experiencing interference from nearby closed access femtocells that are within the critical distance but cannot be connected to \cite{XiaCha10}.  In such a case the SINR would be greatly over-estimated with a standard path loss model, and the gain from interference avoidance or cancellation techniques greatly underestimated.  Or if a nearby picocell was connected to, while interference mostly originated from more distant BSs, the SINR would be greatly underestimated.   In general, standard path loss models may not paint an accurate picture of what happens as networks densify, which is a key theme that we explore in this paper.

\textbf{Millimeter wave cellular networks.}  
The intriguing possibility of using millimeter waves -- $\lambda \in (1,10)$ mm, \ie carrier frequencies of $30$ to $300$ GHz -- for cellular communication makes the revisitation of propagation models particularly urgent \cite{PiKha11, Rap2013ItWillWork}.   A key feature of millimeter wave systems is their sensitivity to blocking \cite{RanRapErk14, BaiHeath2015}.   One recently proposed model with considerable empirical support is to use one path loss exponent $\alpha_{0} \approx 2$ for line of sight (LOS) links and another $\alpha_{1} \approx 3.5$ for non line of sight (NLOS) links \cite{GhoshJSAC}.  Statistically, LOS links are shorter than NLOS links, and a critical distance $R_c$ can be used to approximate the two regimes \cite{SinAnd14b,BaiHeath2015}. Here, $R_c$ is an environmentally dependent random variable, but it could be approximated by the mean LOS distance. For example, in urban parts of New York City and Chicago, this method will lead to $R_c \approx 70$~m \cite{SinAnd14b}, whereas in environments with less blocking $R_c$ would increase.  Although such approximations require considerable further investigation, generalizing to at least a dual-slope model appears essential for millimeter wave cellular systems.


The above examples make clear that a dual (or more) slope path loss model is highly desirable for analysis.  And indeed, such a model is very close to many scenarios in the WINNER II path loss model adopted for 3GPP-LTE standardization \cite{LTERelease9,KyostiMeinilaHentilaEtAl2007} and is well-supported by many measurements see, \cite{ErcegGhassemzadehTaylorEtAl1992,FeuersteinBlackardRappaportEtAl1994a,HamptonMerhebLainEtAl2006} and the references therein.
A more detailed comparison between the dual (or more) slope path loss models
and the models in the standardization activities can be found in \cite{LiuXiaoSoong2014}.

\subsection{Contributions}

The overall contribution of this paper is to analyze the coverage probability (SINR distribution) and potential throughput of a downlink cellular network under multi-slope path loss models, with a focus on the dual-slope model.  This can largely be viewed as a generalization of  \cite{net:Andrews12tcom} which used the standard path loss model and derived fairly simple closed form solutions particularly for the case of $\alpha = 4$.  One notable observation from \cite{net:Andrews12tcom} was that the coverage probability (SINR) can only increase with BS density, and does not depend on the BS density once it is sufficiently large (thus rendering noise negligible compared to interference).  We term this property \emph{SINR invariance}, and will see that it does not even approximately hold with a dual-slope model.

Although the results for a multi-slope path loss model are predictably more complicated than with the standard path loss model, in various limiting cases several crisp statements can be made. Below, we summarize the main contributions of this paper:

\begin{itemize}
\item
We derive numerically tractable integral-form expressions as well as tight closed-form estimates for the coverage probability (\SINR distribution) in cellular networks with multi-slope path loss functions and Poisson distributed BSs.
\item
Focusing on the dual-slope case, we prove that network SIR decreases with increasing network density. Since SNR always increases with network densification, the spectral efficiency is a non-monotonic function of network density, rendering an (finite) optimal density that maximizes the coverage probability, \ie the probability of meeting a particular $\SINR$ target. These results stand in sharp contrast to the SINR invariance observed under the standard path loss model \cite{net:Andrews12tcom, net:Dhillon12jsac}.
\item
However, the network potential throughput still (asymptotically) \emph{linearly} scales with network density $\lambda$ if the near-field path loss exponent $\alpha_0$ is larger than 2. The scaling rate becomes $\Omega(\lambda^{2-\frac{2}{\alpha_0}})$, \ie sublinear, for $1<\alpha_0<2$.
\item
On the other hand, with ultra-densification, \ie the network density $\lambda$ goes to infinity, the potential throughput always scales unboundedly if $\alpha_0>1$, despite the fact that the coverage probability goes to zero in the limit when $\alpha_0<2$. A \emph{phase-transition} happens at $\alpha_0 = 1$: if $\alpha_0 < 1$, ultra-densification ($\lambda \to 0$) always leads ultimately to zero throughput.
\item
The above results are shown to generalize to multi-slope path loss functions, with an arbitrary number of increasing path loss exponents.
\end{itemize}

\subsection{Paper Organization}

In the rest of the paper, we will first introduce the path loss and network models in Section~\ref{sec:sys}. General coverage probability expressions under the dual-slope path loss model are derived in Section~\ref{sec:GenCovProbE}. Section~\ref{sec:SIR} specializes in the interference-limited case, derives key differential properties of the network performance and contrasts them with those found under the standard path loss function. The case with noise is analyzed in Section~\ref{sec:SINRnTPScaling}. Section~\ref{sec:Mslope} generalizes the conclusions drawn in the dual-slope case to the multi-slope case. Concluding remarks are given in Section~\ref{sec:conclu}.


\section{System Model\label{sec:sys}}

\subsection{Network and Path Loss Models}

Consider a typical downlink UE located at the origin $o$.
We assume the BS distribution is governed by a marked Poisson point process (PPP) $\hat\Phi = \{(x_i, h_{x_i})\}\subset \R^2\times \R^+$,
where the ground process $\Phi = \{x_i\}\subset \R^2$ is a homogeneous PPP
with intensity $\lambda$ and $h_{x_i}$ is the (power) fading gain from the BS at $x_i$ to the typical user $o$.\footnote{
The PPP-based cellular network model is well accepted for cellular network analysis, see, \eg \cite{SinAnd14b,Zhang14twc,net:Dhillon12jsac,Bar13Lattice}, and is supported both empirically \cite{net:Andrews12tcom} and theoretically \cite{Bar13Lattice,KeelerRossXia2014}.}
For simplicity but without loss of generality, we assume that all BSs transmit with unit power
and let $l:\R^+\to\R^+$ denote the path loss function.
Then, at the origin $o$, the received power from the BS $x$ is $h_x l(\|x\|)$,
where $\|x-y\|$ is the Euclidean distance between $x$ and $y$.
While some of the results in this paper hold irrespective of the fading distribution,
we will focus on Rayleigh fading, \ie $h_x$ are iid exponentially distributed with unit mean. 
With slight abuse of notation, we may write $l(\|x\|)$ as $l(x)$ for simplicity.

In the following, we (formally) define the few path loss functions of interest.

\begin{definition}[Standard Path Loss Function]
The standard (power-law) path loss function is
\begin{equation}
	l_\textnormal{1}(\alpha;x) = \|x\|^{-\alpha}.
\label{equ:stdplfunc}
\end{equation}
\end{definition}
This simple version is a suitable simplification of \eqref{eq:PLintro} since $K_0$ is assumed to be the same for all links and can simply be folded into the noise power.  Thus, we can write \eqref{equ:stdplfunc} in dB as $- 10 \alpha \log_{10} \| x \|$ and the standard path loss function can be also interpreted as the \emph{single-slope} model
with slope $-\alpha$.  The subscript $1$ indicates this single-slope property.

Many efforts have been made to identify the ``right'' path loss exponent $\alpha$.
It is empirically observed that $\alpha$ is generally best approximated as a constant between $2$ to $5$ which depends on the carrier frequency as well as the physical environment (indoor/outdoor). However, as we noted at the outset, this model has severe limitations, and several motivating examples lead us to consider a dual-slope path loss function.

\begin{definition}[Dual-slope Path Loss Function]
The dual-slope (power-law) path loss function \cite{SarkarJiKimMedouriSalazar-Palma2003} is
\begin{equation}
   l_2(\alpha_0,\alpha_1;x) = \left\{
     \begin{array}{lr}
       \|x\|^{-\alpha_0}, &  \|x\|\leq R_c\\
       \eta \|x\|^{-\alpha_1}, &  \|x\|> R_c,
     \end{array}
   \right.
\label{equ:biexpPL}
\end{equation}
where $\eta \triangleq R_c^{\alpha_1-\alpha_0}$, $R_c>0$ is the \emph{critical distance}, and
$\alpha_0$ and $\alpha_1$ are the \emph{near-} and \emph{far-field path loss exponents} with $0\leq \alpha_0\leq\alpha_1$.
\end{definition}

Clearly, the dual-slope path loss model has two slopes in a dB scale, 
which we stress with the subscript $2$.  The constant $\eta$ is introduced to maintain continuity and complies with the definitions in \cite{SarkarJiKimMedouriSalazar-Palma2003,GoldsmithBook,SchillingMilsteinPickholtzBrunoKanterakisKullbackErcegBiedermanFishmanSalerno1991}.

\begin{remark}
The dual-slope path loss function as defined above is a more general version of the standard path loss function
with the following three important special cases.
\begin{itemize}
\item
The standard path loss function can be retrieved (from the dual-slope path loss function) by setting $\alpha_0=\alpha_1 = \alpha$ in \eqref{equ:biexpPL}.
\item
Letting $R_c \to \infty$, we have $l_2(\alpha_0,\alpha_1;x) = \|x\|^{-\alpha_0}= l_1(\alpha_0;x) $.
Analogously, when $R_c\to 0$, $l_2(\alpha_0,\alpha_1;x) = \eta \|x\|^{-\alpha_1} = \eta l_1(\alpha_1;x) $.
These two special limiting instances of the dual-slope path loss
function will become important in our later coverage analyses.
\item
If $\alpha_0 = 0,~ \alpha_1>2$,
the dual-slope path loss model can be rewritten as
\begin{equation}
	l_2(0,\alpha_1;x) = \min\{1, \eta \|x\|^{-\alpha_1}\},
\label{equ:bdedPL}
\end{equation}
where $\eta = R_c^{\alpha_1}$.
Here, \eqref{equ:bdedPL} can be interpreted as the \emph{bounded} single-slope path loss function,
\ie a fixed path loss up to $R_c$ and a single path loss exponent afterwards.
Indeed, this path loss function is also often used in the literature, see \eg \cite{HeathKountourisBai2013,net:Haenggi11cl,net:Dousse04,FranceschettiDousseTseEtAl2007},
and with experimental support seen in \cite{net:Haenggi05commag}. 
\end{itemize}
\end{remark}


More generally, one can consider a finite number of path loss exponents (and critical distances) and 
obtain a continuous, multi-slope path loss function, which we now define.

\begin{definition}[Multi-slope ($N$-slope) Path Loss Model]
For $N\in\mathbb{N}^+$, the $N$-slope path loss model
\begin{equation}
	l_N(\{\alpha_i\}_{i=1}^{N-1};x) = K_n \|x\|^{-\alpha_n}  
\end{equation}
for $\|x\|\in [R_{n}, R_{n+1}), n\in[N-1]\cup\{0\}$,\footnote{We use $[n]$ to denote the set $\{1,2,\cdots,n\}$.}
where $K_0 = 1$ and $K_n = \prod_{i=1}^n R_i^{\alpha_i-\alpha_{i-1}},~\forall n\in [N-1]$, $0=R_0<R_1<\cdots<R_N=\infty$,
$0\leq \alpha_0\leq\alpha_1\leq\cdots\leq\alpha_{N-1}$, $\alpha_{N-1}>2$.
\label{def:Multi-slopePLM}
\end{definition}

Def.~\ref{def:Multi-slopePLM} is consistent with the piece-wise linear model in \cite[Sect. 2.5.4]{GoldsmithBook}.
Clearly, when $N=2$, the multi-slope path loss function becomes dual-slope,
and $N=1$ gives the standard path loss model.
More importantly, the $N$-slope path loss function provides a means
to study more general path loss functions which decay faster
than power law functions.

For notational simplicity, the path loss exponents parameterizing
the path loss functions may be omitted when they are obvious from the context,
\ie we may write $l_1(\alpha; \cdot)$, $l_2(\alpha_0,\alpha_1; \cdot)$, $l_N(\{\alpha_i\}_{i=1}^{N-1};\cdot)$
as $l_1(\cdot)$, $l_2(\cdot)$, $l_N(\cdot)$, respectively.

%
%

%
%

\subsection{SINR-based Coverage\label{subsect:SINRCov}}

The main metric of this paper is the coverage probability of the typical user at $o$,\footnote{
By the stationarity of $\Phi$, the result will not change if an \emph{arbitrary} location (independent of $\Phi$) is chosen instead of $o$.}
defined as the probability that the received SIR or SINR at the user is larger than
a target $ T$.
When the user is always associated with the nearest BS, 
\ie one with the least path loss and highest average received power,
the SINR can be written as
\begin{equation*}
		\SINR_l = \frac{h_{x^*}l(x^*)}{\sum_{y\in\Phi\setminus\{x^*\}} h_y l(y) +  \sigma^2},
\end{equation*}
where the subscript $l$ in SINR is to emphasize that the SINR is defined
under any path loss function $l$, $x_{*}\triangleq \argmax_{x} l(x)$ and $ \sigma^2$ can be considered as the receiver-side noise
power normalized by the transmit power and other propagation constants, \eg loss at $\|x\|=1$.

Then, the SINR coverage probability can be formalized as
\begin{equation}
	\Pc^\SINR_{l} (\lambda,  T) \triangleq \P(\SINR_l> T).
\label{equ:PcSINRdef}
\end{equation}
where the parameters $(\lambda,  T)$ may be omitted if they are obvious in the context.
It is clear from \eqref{equ:PcSINRdef} that
$\Pc^\SINR_{l} (\lambda,  \cdot)$ is the ccdf of the SINR at the typical user.

In addition to the coverage probability, we further define the coverage density and the potential throughput as 
our primary metrics for the area spectral efficiency under network scaling.
\begin{definition}[Coverage Density and Potential (Single-rate) Throughput]
\label{def:PT}
The \emph{coverage density} of a cellular network under path loss function $l(\cdot)$ is
\begin{equation*}
	\mu_{l}(\lambda, T) \triangleq \lambda \Pc_l^\SINR (\lambda, T)
\end{equation*}
where $\lambda$ is the network (infrastructure) density and $T$ is the $\SINR$ target.  It has units of BSs/Area.

The \emph{potential throughput} is
\begin{equation*}
	\tau_l(\lambda, T) \triangleq \log_2(1+T) \mu_{l}(\lambda, T), 
\end{equation*}
which has units of bps/Hz/m$^2$, the same as area spectral efficiency.
\end{definition}

Whereas the coverage probability can be used to capture the spectral efficiency distribution (since they have a 1:1 relation),
the potential throughput gives an indication of the (maximum) cell splitting gain, which would occur if all BSs remain fully loaded as the network densifies.
To see this, consider the case of \emph{simultaneous densification},
\ie the densities of network infrastructure (BSs) and users scale at the same rate.
Assuming the user process is stationary and independent of $\Phi$,
it is not difficult to observe that
under simultaneous densification, the scaling of the area spectral efficiency (ASE), defined as the number
of bits received per unit time, frequency and area, is
\emph{the same} as the scaling of the potential throughput\footnote{This argument could be made rigorous by introducing further assumptions on the scheduling procedure and traffic statistics but is beyond the content of this paper.},
which by the Def.~\ref{def:PT} also equals that of the coverage density.

%
%
%
%

For example, using the standard path loss function,
\cite{net:Andrews12tcom} shows that the network density does not change the SINR distribution
in the interference-limited case, which leads to the potential throughput growing
linearly with the network density with simultaneous densification and implies a linear scaling for the cellular network area spectral efficiency.

\section{The General Coverage Probability Expressions\label{sec:GenCovProbE}}

The stochastic geometry framework provides a tractable way to characterize the coverage probability for cellular networks. Generally speaking, an integral form of the coverage probability can be derived under arbitrary fading regardless of the path loss function \cite{net:Sousa90,net:Haenggi09jsac}. However, Rayleigh fading (having an exponential power pdf) is nearly always used due to its outstanding tractability, and as seen in \cite{net:Andrews12tcom}, it yields similar results to other fading/shadowing distributions (as long as they have the same mean) due to the spatial averaging inherent to stochastic geometry.  Admittedly, the standard path loss function does give some extra tractability which cannot be duplicated with more general path loss functions.

In this section, we give an explicit expression for the coverage probability
for general path loss function and demonstrate that it can be simplified in terms
of Gauss hypergeometric functions under the dual-slope power law path loss function.

\begin{lemma}
The coverage probability under a nearest BS association
and general path loss function $l(\cdot)$ is $\Pc^\SINR_l(\lambda, T) =$
\begin{multline}
\lambda\pi \int_0^\infty \exp\left({-\lambda\pi y \Big(1+\int_1^\infty 
									\frac{ T}{ T+\frac{l(\sqrt{y})}{l(\sqrt{t y})}}\d t\Big)}\right) \\
									 \times e^{- T  \sigma^2/l(\sqrt{y})} \d y.
\label{equ:NBSACvrgP}
\end{multline}
\label{lem:CovPGenPL}
\end{lemma}


The proof of Lemma~\ref{lem:CovPGenPL} is analogous to that of \cite[Theorem 1]{net:Andrews12tcom}.
It is a result of the probability generating functional (PGFL) and the nearest neighbor distribution of the PPP,
and changes of variables.
We omit the proof for brevity.
Note that the tractability exposed in \cite{net:Andrews12tcom} hinges on the fact that
$\frac{l(\sqrt{y})}{l(\sqrt{ty})}$ is independent of $y$ under the standard path loss function
which does not apply for general path loss functions.
However, \eqref{equ:NBSACvrgP} does allow numerical computation for the coverage
probability for general path loss functions.
For the dual-slope path loss function, 
\eqref{equ:NBSACvrgP} can be further simplified as in the following theorem.

\begin{theorem}
The coverage probability under the dual-slope path loss function
is
\begin{multline}
\Pc^\SINR_{l_2}(\lambda, T) =
				\lambda\pi R_c^2  \int_0^1 e^{-\lambda\pi R_c^2 I(\delta_0,\delta_1, T; x) 
				- T  \sigma^2 x^{\frac{\alpha_0}{2}} R_c^{\alpha_0}} \d x \\
			+ \lambda\pi R_c^2 \int_1^{\infty} e^{ -\lambda\pi R_c^2 x C_{-\delta_1}( T) 
			- T  \sigma^2 x^{\frac{\alpha_1}{2}} R_c^{\alpha_0}} \d x  ,
\label{equ:2expCP}
\end{multline}
where $I(\delta_0,\delta_1, T; x) =$
\[			C_{\delta_0}\left(\frac{1}{ T x^{\frac{1}{\delta_0}}}\right)
				+C_{-\delta_1}({ T x^{\frac{1}{\delta_0}}})
				+ x \left(1-C_{\delta_0}\left(\frac{1}{ T}\right)\right) -1,
\]
 $C_\beta(x) = {_2F}_1(1,\beta;1+\beta;-x)$, where
${_2F}_1 (a,b;c;z)$ is the \emph{Gauss hypergeometric function}, $\delta_0=2/\alpha_0$,\footnote{If $\alpha_0=0$, we interpret $\delta_0=\infty$.} $\delta_1=2/\alpha_1$.\label{thm:2exp}
\end{theorem}

\begin{IEEEproof}
The proof follows directly from Lemma~\ref{lem:CovPGenPL} and changes of variables.
\end{IEEEproof}

The first term in \eqref{equ:2expCP} represents the coverage probability when the distance to the serving BS is less than the critical distance $R_c$, and the second term is the coverage probability when it is farther than $R_c$. The intervals of integral $(0,1)$ and $(1,\infty)$ result from a change of variables.

In most reasonably dense (e.g. urban) existing cellular networks, interference dominates the noise power,
making the signal-to-interference ratio (SIR), $\SIR_l \triangleq \SINR_l |_{ \sigma^2=0}$, an accurate approximation to SINR.
Such an approximation has been adopted in many cellular network analyses, see \eg \cite{Zhang14twc}.
If we define the SIR coverage probability
$\Pc^\SIR_{l} (\lambda,  T) \triangleq \P(\SIR_l> T)$ as the probability that the received SIR
at the typical user is above the threshold $ T$,
Theorem~\ref{thm:2exp} yields the following important observation.

\begin{fact}[Near-field-BS Invariance]
For two dual-slope path loss function $l_2(\cdot)$ and $l'_2(\cdot)$ with
the same path loss exponents but different critical distances $R_c$ and $R'_c$,
the effect of density and the critical distance
on the SIR coverage probability is \emph{equivalent}
in the sense that $\Pc_{l_2}^\SIR(\lambda, T) = \Pc_{l_2'}^\SIR(\lambda', T)$
as long as $\lambda R_c^2 = \lambda' (R'_c)^2$, \ie
the mean numbers of the near-field BSs are the same.
\label{fact:lambdaR2}
\end{fact}

\begin{remark}[Loss of SIR-invariance]
Under the standard path loss model,
the $\SIR$ coverage probability is independent of the network density \cite{net:Andrews12tcom}, i.e. \SIR-invariance holds.
Fact~\ref{fact:lambdaR2} looks similar but is much weaker than the \SIR-invariance property.
Under the dual-slope path loss function, $\Pc_{l_2}^\SIR(\lambda,R_c)$ is held constant only if
$R_c$ scales with $1/\sqrt{\lambda}$ as the network densifies.
Since empirically $R_c\propto f_\textnormal{c}$,
the ambition of maintaining the same spectral efficiency with higher network density is equivalent to asking for more bandwidth at the lower end of the spectrum, an unrealistic request.
\end{remark}


\begin{remark}[Requirements for Finite Interference]
Unlike for the standard path loss function, where $\alpha>2$ is typically required to
guarantee (almost surely) bounded interference,
Theorem~\ref{thm:2exp} and \eqref{equ:2expCP} only requires $\alpha_1>2$.
Intuitively, the interfering region under $\alpha_0$ is always finite and thus does not contribute
infinite interference (at finite network density) and $\alpha_1>2$ guarantees the interference from beyond the critical distance is
bounded.
\end{remark}

\begin{remark}[Simplifying Special Cases]
For some particular choices of $\alpha_0$ and $\alpha_1$,
the need for hypergeometric functions in \eqref{equ:2expCP}
can be eliminated.
In particular, we have
$C_1(x) = \frac{\log(1+x)}{x}$, $C_{-\frac{1}{2}}(x)=1+\sqrt{x}\arctan \sqrt{x}$, $C_\frac{1}{2}(x)=\frac{\arctan\sqrt{x}}{\sqrt{x}}$, $C_2(x) = \frac{2(x-\log(1+x))}{x^2}$ and $C_\infty(x) = \frac{1}{1+x}$.
Consequently, many important special cases can be expressed without special functions
including $[\alpha_0\; \alpha_1] = [2\; 4],[1\; 4],[0\; 4]$.
\end{remark}

Among these several special cases, the most interesting one is probably $[\alpha_0\; \alpha_1] = [2\; 4]$
which coincides with the well-known two-ray model.  We thus conclude this section with a corollary highlighting this case.

\begin{corollary}
The \SINR coverage probability under a dual-slope path loss function with $\alpha_0 = 2$, $\alpha_1 = 4$ is
\begin{multline}
\Pc^\SINR_{l_2}(\lambda, T) = 
				\lambda\pi R_c^2 \int_0^1 e^{-\lambda\pi R_c^2 I(\delta_0,\delta_1, T; x) 
				- T  \sigma^2 x R_c^2} \d x \\
			+ \lambda\pi R_c^2 \int_1^{\infty} e^{ -\lambda\pi R_c^2 x (1+\sqrt{T}\arctan\sqrt{T}) 
			- T  \sigma^2 x^{2} R_c^2} \d x  ,
\end{multline}
where
\begin{multline*}		
I(\delta_0,\delta_1, T; x) =xT\log\left(1+\frac{1}{xT}\right) \\
+ \sqrt{xT}\arctan\sqrt{xT} + x\left(1-T\log\big(1+\frac{1}{T}\big)\right).
\end{multline*}
\end{corollary}


\section{The Interference-limited Case\label{sec:SIR}}

Theorem~\ref{thm:2exp} gives an exact expression of the coverage probability in the cellular
network modeled by a PPP.
In this section, we refine our understanding about the dual-slope path loss function by
comparing against the standard path loss function, and highlighting the differences.

\begin{lemma}
For an arbitrary marked point pattern (including fading) $\hat\Phi (\omega) \subset \R^2\times \R^+$
associated with sample $\omega\in\Omega$
and any $ T>0$,
\begin{itemize}
\item
$\SIR_{l_1(\alpha_0;\cdot)} (\omega) >  T$ implies $\SIR_{l_2(\alpha_0,\alpha_1;\cdot)} (\omega) \geq  T$,
\item
$\SIR_{l_2(\alpha_0,\alpha_1;\cdot)} (\omega) >  T$ implies $\SIR_{l_1(\alpha_1;\cdot)} (\omega) \geq  T$,
\end{itemize}
where $\SIR_l(\omega)$ is the $\SIR$ at the typical user under path loss function $l(\cdot)$ for a sample $\omega$.
\label{lem:SIRbounds}
\end{lemma}

\begin{IEEEproof}
See Appendix~\ref{app:SIRbounds}.
\end{IEEEproof}

\begin{remark}[Generality of SIR Bounds]
Lemma~\ref{lem:SIRbounds} is stated for an arbitrary realization of the network topology
and fading, and does not depend on any statistical assumptions.
It is purely based on the nature of the path loss functions in question.
\end{remark}

An immediate consequence of Lemma~\ref{lem:SIRbounds} is the SIR coverage ordering
of cellular networks with general fading and BS location statistics.

\begin{theorem}\label{thm:PcSIRordering}
For random wireless networks modeled by arbitrary point process and fading, under the nearest BS
association policy, the following $\SIR$ coverage probability ordering holds for arbitrary $0\leq\alpha_0\leq \alpha_1$:
\begin{equation*}
	\Pc^{\SIR}_{l_1(\alpha_1;\cdot)}(\cdot, T)\geq \Pc^{\SIR}_{l_2(\alpha_0,\alpha_1;\cdot)}(\lambda, T) \geq \Pc^{\SIR}_{l_1(\alpha_0;\cdot)}(\cdot, T).
\end{equation*}
\end{theorem}
The proof of Theorem~\ref{thm:PcSIRordering} follows directly from Lemma~\ref{lem:SIRbounds}
and is omitted from the paper.

In addition to being an important characterization of the dual-slope
path loss function, Theorem~\ref{thm:PcSIRordering} leads to the following interesting
fact that has been observed in many special cases. 

\begin{corollary}
For random wireless networks modeled by arbitrary point process and fading, under the nearest BS
association policy, the $\SIR$ coverage probability
is a monotonically increasing function of the path loss exponent for the standard path loss function.
\label{cor:SICcovSinglePLEmonotonicity}
\end{corollary}

Cor.~\ref{cor:SICcovSinglePLEmonotonicity} follows directly from the observation
that $\Pc^{\SIR}_{l_1(\alpha_1;\cdot)} (\cdot, T) \geq \Pc^{\SIR}_{l_1(\alpha_0;\cdot)} (\cdot, T)$ is true for all $0\leq\alpha_0\leq \alpha_1$
(including the case of $\alpha_1\leq 2$).
Although given Rayleigh fading and the PPP model,
the fact that SIR coverage monotonically increases
with $\alpha$ is known before, \eg easily inferred from the expressions in \cite{net:Andrews12tcom}.
Theorem~\ref{thm:PcSIRordering} shows that such monotonicity is a nature of the (standard) path loss function
and is independent of any network and fading statistics.
Furthermore, the theorem includes the case $\alpha\leq 2$, which was often excluded in conventional analyses.


Since the coverage probability under the standard path loss function is well-known for $\alpha>2$ \cite{net:Andrews12tcom}, Theorem~\ref{thm:PcSIRordering} leads to computable bounds on the SIR coverage probability with the dual-slope path loss function.
A natural question follows: what if the dual-slope model is applied but with $\alpha_0\leq 2$?
Although $\alpha\leq 2$ is not particularly interesting under the standard path loss function since it is both intractable and not empirically supported, a
small ($\leq 2$) near-field path loss exponent is relevant under the dual-slope model since
both early reports in the traditional cellular frequency bands \cite{SchillingMilsteinPickholtzBrunoKanterakisKullbackErcegBiedermanFishmanSalerno1991}
and recent measurements at the millimeter wave bands \cite{AzarRap2013ICC28Goutdoor} suggest that small near-field path loss exponents are definitely plausible.  Intuitively, a small $\alpha_0 < 2$ simply means that for short distances, the path loss effects are fairly negligible versus for example the positive impact of reflections or directionality. The following proposition highlights an interesting feature of this small $\alpha_0$ case. 

\begin{proposition}
Under the dual-slope pathloss model, when $\alpha_0\leq 2$,
the $\SIR$ and $\SINR$ coverage probabilities $\Pc^\SIR_{l_2}$ and $\Pc^\SINR_{l_2}$ 
go to zero as $\lambda\to\infty$.
\label{prop:densezerocoverage}
\end{proposition}

\begin{IEEEproof}
See Appendix~\ref{app:densezerocoverage}.
\end{IEEEproof}

The most important implication from Prop.~\ref{prop:densezerocoverage} is that
ultra-densification could eventually lead to near-universal outage if $\alpha_0\leq 2$.
It is worth stressing that this asymptotically zero coverage probability happens if and only if $\alpha_0\leq 2$ and for any $\alpha_0>2$,
(still) $\Pc^\SINR_{l_2(\alpha_0,\alpha_1;\cdot)}(\lambda, T)>0$ for all $ T,\lambda>0$.

Combining Prop.~\ref{prop:densezerocoverage} with Theorem~\ref{thm:PcSIRordering} leads to the following corollary.

\begin{corollary}
Under the standard path loss model, the typical user has an SINR and SIR coverage probability of zero almost surely
if the path loss exponent is no larger than $2$.
\label{cor:0CovP}
\end{corollary}


The following lemma strengthens Theorem~\ref{thm:PcSIRordering} by showing that
the upper and lower bounds on $\Pc^{\SIR}_{l_2(\alpha_0,\alpha_1;\cdot)}$
are achievable by varying the network density.

\begin{lemma}
The following is true for any $ T\geq 0$:
\begin{itemize}
\item
$\lim_{\lambda\to\infty} \Pc^\SIR_{l_2(\alpha_0,\alpha_1;\cdot)} (\lambda,  T) = \lim_{\lambda\to\infty} \Pc^\SINR_{l_2(\alpha_0,\alpha_1;\cdot)}(\lambda, T) = \Pc^\SIR_{l_1(\alpha_0;\cdot)} (\cdot,  T)$.
\item
$\lim_{\lambda\to 0} \Pc^\SIR_{l_2(\alpha_0,\alpha_1;\cdot)}(\lambda,  T) = \Pc^\SIR_{l_1(\alpha_1;\cdot)} (\cdot,  T)$
\end{itemize}
\label{lem:BoundAchi}
\end{lemma}

\begin{IEEEproof}
First, we realize that both $\Pc^\SIR_{l_1(\alpha_0;\cdot)}(\lambda,  T)$ and $\Pc^\SIR_{l_1(\alpha_1;\cdot)}(\lambda,  T)$
are independent from $\lambda$.
This fact is most well known for the case $\alpha_0,\alpha_1 > 2$, see, \eg \cite{net:Andrews12tcom}, but Cor.~\ref{cor:0CovP}
confirms that it is true for all $\alpha_0,\alpha_1>0$.
By Theorem~\ref{thm:PcSIRordering}, we have $\Pc^{\SIR}_{l_1(\alpha_1;\cdot)}(\cdot, T)\geq \Pc^{\SIR}_{l_2(\alpha_0,\alpha_1;\cdot)}(\lambda, T) \geq \Pc^{\SIR}_{l_1(\alpha_0;\cdot)}(\cdot, T)$ for all $\lambda>0$.
To show the convergence, we make use of Fact~\ref{fact:lambdaR2}.
Instead of letting $\lambda\to\infty$ (and $\lambda\to 0$ resp.), we consider equivalently
$R_c\to\infty$ ($R_c\to 0$ resp.).
But by the definition of the dual-slope path loss function, such
scaling results in $l_1(\alpha_0;\cdot)$ ( $\eta l_1(\alpha_1;\cdot)$ resp.).
The lemma is completed by observing that $ \Pc^\SIR_{l_2} \to \Pc^\SINR_{l_2}$ as $\lambda\to\infty$.
\end{IEEEproof}

Theorem~\ref{thm:PcSIRordering} and Lemma~\ref{lem:BoundAchi} point to the perhaps counter-intuitive
conclusion that $\SIR$ coverage probability decays with network densification.
This is formalized in the following lemma.

\begin{lemma}[SIR monotonicity]
Under the dual-slope path loss function and arbitrary fading distribution, $ \Pc^\SIR_{l_2} (\lambda_1, T) \geq \Pc^\SIR_{l_2}(\lambda_2, T)$, for all $\lambda_1\leq\lambda_2$, $ T \geq 0$ and $0\leq\alpha_0\leq\alpha_1$.
\label{lem:SIRmono}
\end{lemma}

\begin{IEEEproof}
See Appendix~\ref{app:SIRmono}.
\end{IEEEproof}

\begin{figure}[tbh]
\centering
\psfrag{lambda}[c][c]{$\lambda$}
\psfrag{Pc}[c][t]{$\Pc^\SIR_{l_2(\alpha_0,\alpha_1;\cdot)}$}
\begin{overpic}[width=\linewidth]{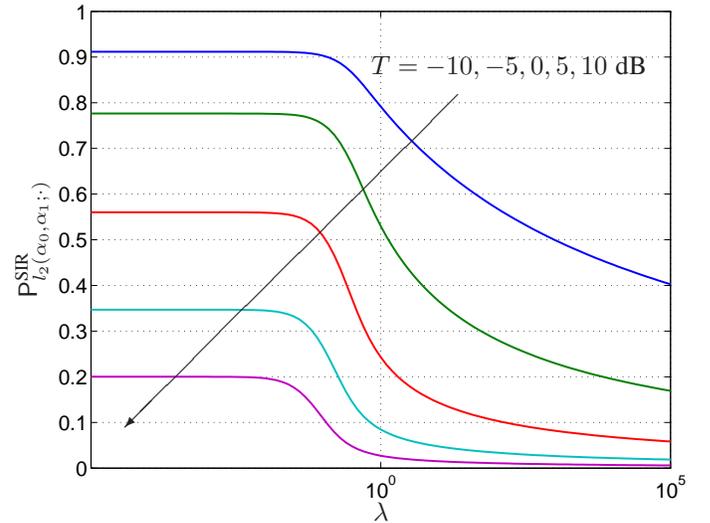}
	\put(65,65){\vector(-1,-1){50}}
	\put(52,68){$ T=-10,-5,0,5,10$ dB}
\end{overpic}
\caption{SIR coverage scaling with network density when $\alpha_0 = 2$, $\alpha_1 = 4$, $R_c=1$. \label{fig:SIRcvrgProbxlambda-10-10_alpha0_2_alpha1_4}}
\end{figure}

Fig.~\ref{fig:SIRcvrgProbxlambda-10-10_alpha0_3_alpha1_4} plots the SIR coverage probability as a function of $\lambda$
for $ T = -10,-5,0,5,10$ dB (top to bottom).
Consistent with Lemma~\ref{lem:SIRmono}, we see SIR coverage decreases with increasing density.
The convergence of $\Pc^\SIR_{l_2} (\lambda, T)$ as $\lambda\to\infty$
and $\lambda\to 0$ is also verified in the figure.
In Fig.~\ref{fig:SIRcvrgProbxlambda-10-10_alpha0_3_alpha1_4}, we use $\alpha_0 = 3>2$ and thus positive
coverage probability is expected as $\lambda\to\infty$.
In contrast, Fig.~\ref{fig:SIRcvrgProbxlambda-10-10_alpha0_2_alpha1_4}
demonstrates the coverage probability scaling predicted by Prop.~\ref{prop:densezerocoverage} and Lemma~\ref{lem:SIRmono},
\ie $\Pc^\SIR_{l_2} (\lambda, T)$
keeps decreasing (to zero) regardless of $ T$ as $\lambda$ increases.
The sharp visual difference between Figs.~\ref{fig:SIRcvrgProbxlambda-10-10_alpha0_3_alpha1_4}~and~\ref{fig:SIRcvrgProbxlambda-10-10_alpha0_2_alpha1_4}
highlights the \emph{phase transition} on $\alpha_0$ with 2 being the critical exponent. 

\begin{figure}[th]
\centering
\psfrag{lambda}[c][c]{$\lambda$}
\psfrag{Pc}[c][t]{$\Pc^\SIR_{l_2(\alpha_0,\alpha_1;\cdot)}$}
\begin{overpic}[width=\linewidth]{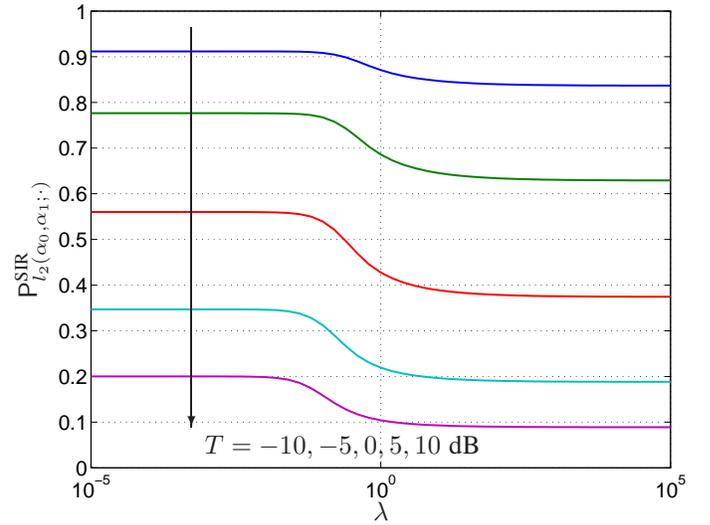}
	\put(25,75){\vector(0,-1){60}}
	\put(27,11){$ T=-10,-5,0,5,10$ dB}
\end{overpic}
\caption{SIR coverage scaling with network density when $\alpha_0 = 3$, $\alpha_1 = 4$, $R_c=1$. \label{fig:SIRcvrgProbxlambda-10-10_alpha0_3_alpha1_4}}
\end{figure}

Intuitively, one can understand this result by considering the best-case scenario for SIR which would occur at a low density, where the UE is connected to the nearest BS which is in the near-field and all the interfering BSs are located in the far-field and thus more rapidly attenuate.  Increasing the density in such a case could only reduce SIR, since interfering BSs would soon be added in the near-field.  This density regime is where we observe the the transition from higher to lower SIR in 
Figs.~\ref{fig:SIRcvrgProbxlambda-10-10_alpha0_3_alpha1_4}~and~\ref{fig:SIRcvrgProbxlambda-10-10_alpha0_2_alpha1_4}.  Asymptotically, an infinite number of BSs will be present in the near-field, and we are back to SIR-invariance as observed for the standard path loss model since we have only a single relevant path loss exponent, $\alpha_0$.

A hasty conclusion from this discussion is that to optimize the SIR coverage probability, one
can simply let the density of the network go to zero. While this statement is true, it is not of  much practical relevance
since as the network density goes to zero the received signal power goes to zero as well, 
and the network is no longer interference-limited. Thus, unlike the standard path loss case, where the interference-limited
assumption is often justifiable in the coverage analysis,
the dual-slope path loss function increases the importance of including noise.

\section{SINR Coverage and Throughput Scaling\label{sec:SINRnTPScaling}}

\subsection{The Tension between SIR and SNR\label{sec:SIRnSNR}}

As shown in Sect.~\ref{sec:SIR}, BS densification generally reduces
the SIR coverage under the dual-slope path loss model.
Yet, bringing the BSs closer to the users clearly increases SNR.
Thus, the optimal density of the network introduces a tradeoff between SIR and SNR.
While a closed-form expression of the SINR coverage does not exist in general,
there are multiple ways to characterize the coverage probabilty
as a function of the network density in addition to directly applying the integral
expression in Theorem~\ref{thm:2exp}.

\subsubsection{SNR Coverage Analysis\label{subsec:SNR}}

To complement the SIR coverage analysis in Sect.~\ref{sec:SIR},
it is natural to focus on the SNR coverage probability, defined as the probability
that $\SNR_l \triangleq h_{x^*} l(x^*) /  \sigma^2 >  T$.
Such analysis has not attracted much attention under the standard path loss model due to the SINR monotonicity
under the standard path loss function (\ie $\SINR$ increases monotonically with network density),
but becomes relevant for the dual-slope path loss model.
The analysis is also important for noise-limited systems including the emerging mmWave networks \cite{RanRapErk14}.

\begin{lemma}[The SNR Coverage Probability]\label{lem:SNRCovP}
The SNR coverage probability under the dual-slope path loss model is
\begin{multline}
	\Pc^\SNR_{l_2}(\lambda,  T) = \lambda\pi R_c^2 \int^1_0 e^{-\lambda \pi R_c^2 y 
																						-  T  \sigma^2 R_c^{\alpha_0} y^{\frac{\alpha_0}{2}}} \d y		\\
																						+ \lambda \pi R_c^2 \int_1^\infty e^{-\lambda\pi R_c^2 y -  T  \sigma^2 R_c^{\alpha_0} y^{																						\frac{\alpha_1}{2}}}\d y.
\label{equ:SNRCovP}
\end{multline}
\end{lemma}

The proof of Lemma~\ref{lem:SNRCovP} is strightforward, using the well-known distance distribution
in Poisson networks \cite{net:Haenggi05tit} and so is omitted from the paper.
The first term in \eqref{equ:SNRCovP} corresponds to the case where the serving BS station
is within distance $R_c$ to the typical user and the second term to the case where the serving BS
is farther than $R_c$ away from the user.

In general, the SNR coverage probability \eqref{equ:SNRCovP}
cannot be written in closed-form. But for the special case $\alpha_0=2$
and $\alpha_1 = 4$, it can be simplified as in the following corollary.

\begin{corollary}
For $\alpha_0=2$, $\alpha_1 = 4$, the $\SNR$ coverage probability is
\begin{multline}
		\Pc^\SNR_{l_2(2,4;\cdot)}(\lambda,  T) = \frac{\lambda\pi}{\lambda\pi+ T  \sigma^2} (1-e^{-(\lambda\pi+ T  \sigma^2) R_c^2}) 	\\
													+\frac{\lambda \pi^{\frac{3}{2}}R_c}{\sqrt{ T  \sigma^2}} e^{\frac{\lambda^2\pi^2 R_c^2}{4 T  \sigma^2}}
													\Q\left(\frac{\lambda\pi+2 T  \sigma^2}{\sqrt{2 T  \sigma^2}}R_c\right),
\end{multline}
where $\Q(x) = \frac{1}{\sqrt{2\pi}}\int_x^\infty e^{-{t^2}/{2}}\d t$.
\end{corollary}

Naturally, both $\Pc^\SNR_{l_2}(\lambda,  T)$ and $\Pc^\SIR_{l_2}(\lambda,  T)$ are upper bounds on $\Pc^\SINR_{l_2}(\lambda,  T)$ and the former is asymptotically tight for $\lambda\to 0$ and the latter for $\lambda\to\infty$. Taking the minimum of them could result in an informative characterization of the interplay between interference and noise as the network densifies.

\begin{figure}[tbh]
\centering
\psfrag{lambda}[c][c]{$\lambda$}
\psfrag{Pc}[c][t]{$\Pc^\SINR_{l_2}$, $\Pc^\SIR_{l_2}$, $\Pc^\SNR_{l_2}$}
\begin{overpic}[width=\linewidth]{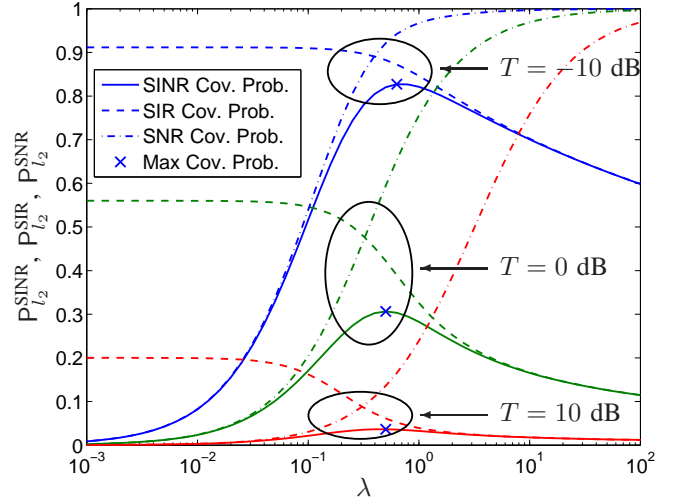}
	\put(74,65){\vector(-1,0){7}}
	\put(76,64){$ T=-10$ dB}
	\put(74,35){\vector(-1,0){10}}
	\put(76,34){$ T=0$ dB}
	\put(74,13){\vector(-1,0){10}}
	\put(76,12){$ T=10$ dB}
\end{overpic}
\caption{SINR, SIR, and SNR coverage scaling vs. network density, with $\alpha_0 = 2$, $\alpha_1 = 4$, $R_c=1$, $ \sigma^2 = 1$.\label{fig:SINRcvrgProbxalpha02_alpha14_lambda1e-05_10000000000}}
\end{figure}

\begin{figure}[tbh]
\centering
\psfrag{lambda}[c][c]{$\lambda$}
\psfrag{Pc}[c][t]{$\Pc^\SINR_{l_2(\alpha_0,\alpha_1;\cdot)}$, $\Pc^\SIR_{l_2(\alpha_0,\alpha_1;\cdot)}$, $\Pc^\SNR_{l_2(\alpha_0,\alpha_1;\cdot)}$}
\begin{overpic}[width=\linewidth]{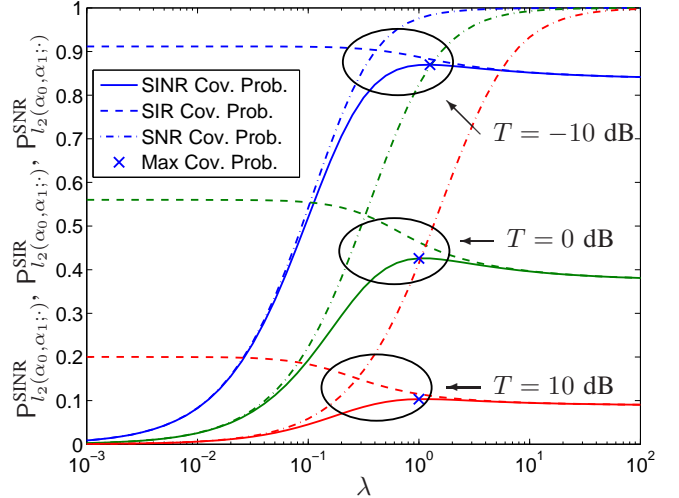}
	\put(73,55){\vector(-1,1){5}}
	\put(75,54){$ T=-10$ dB}
	\put(75,39){\vector(-1,0){5}}
	\put(77,38){$ T=0$ dB}
	\put(73,17){\vector(-1,0){5}}
	\put(75,16){$ T=10$ dB}
\end{overpic}
\caption{SINR, SIR, and SNR coverage scaling vs. network density, with $\alpha_0 = 3$, $\alpha_1 = 4$, $R_c=1$, $ \sigma^2 = 1$.\label{fig:SINRcvrgProbxalpha03_alpha14_lambda0point0001_100}}
\end{figure}

Fig.~\ref{fig:SINRcvrgProbxalpha02_alpha14_lambda1e-05_10000000000} 
and Fig.~\ref{fig:SINRcvrgProbxalpha03_alpha14_lambda0point0001_100}
compare the coverage probability
for the case $\alpha_0=2,3$ and $\alpha_1 = 4$.
As expected, we observe that the SINR coverage probability is maximized for some finite $\lambda$
which effectively strikes a balance between SIR coverage and SNR coverage.
The former decreases with $\lambda$; the latter increases with $\lambda$;
both of them are upper bounds on the SINR coverage probability.
Fig.~\ref{fig:SINRcvrgProbxalpha02_alpha14_lambda1e-05_10000000000} also verifies Prop.~\ref{prop:densezerocoverage}
as the coverage probability goes to zero as $\lambda\to\infty$.
Fig.~\ref{fig:SINRcvrgProbxalpha03_alpha14_lambda0point0001_100} is consistent with Lemma~\ref{lem:BoundAchi} and shows that ultra-densification will lead to constant positive coverage probability if $\alpha_0>2$.
The numerical example also suggests that in this case, the decay of coverage probability with
densification is smaller for low and high SINR but larger for medium SINR. 


\subsubsection{\SINR Distribution for the Two-ray Model}

For the special case of $\alpha_0 = 2$ and $\alpha_1 = 4$, it is possible to derive
a tight lower bound on the coverage probability and thus the ccdf of $\SINR$ as 
an alternative to the numerical integral in Theorem~\ref{thm:2exp}.

\begin{proposition}
For $\alpha_0=2$ and $\alpha_1 =4$, we have the (closed-form) lower bound in \eqref{equ:SINR24LB} (at the top of the next page),
where $\gtrsim$ denotes larger than and asymptotically equal to (with respect to both $\lambda\to 0$ and $T \to 0$),
$C_{-\frac{1}{2}}(x)=1+\sqrt{x}\arctan \sqrt{x}$,
$\rho_0(\lambda, T, \sigma^2) = {\lambda\pi(1+ T)+ T  \sigma^2}$,
$\rho_1(\lambda, T, \sigma^2) = {\lambda\pi R_c}/{\sqrt{ T  \sigma^2}}$,
$\gamma_\textnormal{E} \approx 0.577$ is the Euler-Mascheroni constant, $\Ei(x) = \gamma(0,x) = \int_x^\infty \frac{e^{-t}}{t} \d t$
is the exponential integral function.
\label{prop:SINRcovLB}
\end{proposition}

\begin{figure*}[!t]
\normalsize
\begin{multline}	
	\Pc^\SINR_{l_2(2,4;\cdot)}(\lambda, T) \gtrsim
			\frac{\lambda\pi}{\rho_0(\lambda, T, \sigma^2)}
				\left( 1- e^{ -{\rho_0(\lambda, T, \sigma^2) R_c^2}} \right)
				e^{\frac{\lambda\pi T}{\rho_0(\lambda, T, \sigma^2)}
				\left( 1- \frac{\gamma_\textnormal{E} + \log\left(\rho_0(\lambda, T, \sigma^2)\right) + \Ei(\rho_0(\lambda, T, \sigma^2))}{1-\exp(\rho_0(\lambda, T, \sigma^2) R_c^2)} \right)} \\
			+ \sqrt{\pi}\rho_1(\lambda, T, \sigma^2) e^{\frac{1}{4}\left({C_{-\frac{1}{2}}( T)}\rho_1(\lambda, T, \sigma^2)\right)^2}
				\Q\left(\frac{1}{\sqrt{2}}\rho_1(\lambda, T, \sigma^2) C_{-\frac{1}{2}}( T) + \sqrt{2 T  \sigma^2} R_c\right)
\label{equ:SINR24LB}
\end{multline}
\hrulefill
\vspace*{4pt}
\end{figure*}

\begin{IEEEproof}
See Appendix~\ref{app:SINRcovLB}.
\end{IEEEproof}

\begin{figure}[tbh]
\centering
\psfrag{theta}[c][b]{$ T$}
\psfrag{Pc}[c][t]{$\Pc^\SINR_{l_2(\alpha_0,\alpha_1;\cdot)}$}
\psfrag{ccc}{$\lambda = 0.1$}
\psfrag{bbb}{$\lambda = 10$}
\psfrag{aaa}{$\lambda = 1$}
\begin{overpic}[width=\linewidth]{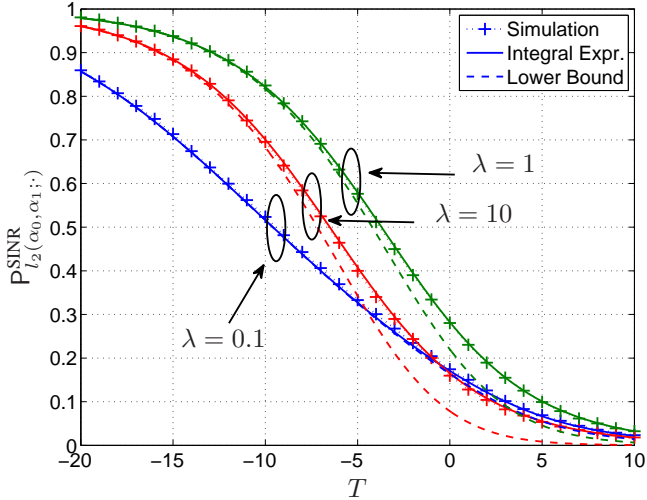}
\end{overpic}
\caption{SINR ccdf from simulation, Theorem~\ref{thm:2exp} and Prop.~\ref{prop:SINRcovLB}.
Here, $\alpha_0 = 2$, $\alpha_1 = 4$, $R_c=1$, $ \sigma^2 = 1$.\label{fig:SINRccdfalpha02_alpha14_lambda0point1_10_c}}
\end{figure}

Prop.~\ref{prop:SINRcovLB} does not involve numerical integral and is instead based
on two well-known special functions: the Q-function and the exponential integral function.
Fig.~\ref{fig:SINRccdfalpha02_alpha14_lambda0point1_10_c} compares the lower bound in Prop.~\ref{prop:SINRcovLB}
with simulation results and the integral expression in Theorem~\ref{thm:2exp}.
The figure numerically verifies the asymptotic tightness of the bound for $ T\to 0$ and/or $\lambda\to 0$.
With the $\lambda = 1$ curve on top,
Fig.~\ref{fig:SINRccdfalpha02_alpha14_lambda0point1_10_c} also confirms that \SINR
does not increase with network density\footnote{More precisely, the \SINR first increases with the network density
(in the noise-limited regime) and then decreases with the network density (in the interference-limited regime).}, as expected.

\subsection{Throughput Scaling}

The coverage probability analysis alone does not provide a complete characterization of the network
performance scaling under densification.
To understand how the area spectral efficiency scales, we further study the potential throughput
defined in Sect.~\ref{subsect:SINRCov}.
By the definition of the potential throughput (Def.~\ref{def:PT}) and Theorem~\ref{thm:PcSIRordering},
we immediately obtain following lemma.

\begin{lemma}
Under the dual-slope path loss model and full-load assumption, the potential network throughput grows \emph{linearly} with BS
density $\lambda$ (as $\lambda\to\infty$) if $\alpha_0>2$.
\label{lem:capacityLinearScaling}
\end{lemma}
\begin{IEEEproof}
If $\alpha_0>2$, $\Pc^\SIR_{l_1(\alpha_1;\cdot)}$ and $\Pc^{\SIR}_{l_1(\alpha_0;\cdot)}$ are positive and invariant
with network density $\lambda$ for any $ T\geq 0$ \cite{net:Andrews12tcom}.
Under the full load assumption, the coverage density $\lambda \Pc^\SIR_{l_1(\alpha_0;\cdot)} \leq \mu_{l_2(\alpha_0,\alpha_1;\cdot)} \leq \lambda \Pc^\SIR_{l_1(\alpha_1;\cdot)}$.
Consequently, $ \mu_{l_2}(\lambda,T)= \Theta(\lambda)$ and in interference-limited network.
By the definition of potential throughput in Def.~\ref{def:PT}, we further have $ \tau_{l_2}(\lambda,T)= \Theta(\lambda)$.
When $\lambda\to\infty$, interference dominates noise and thus the same throughput scaling holds
in noisy networks.
\end{IEEEproof}

While Lemma~\ref{lem:capacityLinearScaling} is encouraging,
it is only for the case $\alpha_0 > 2$.
On the other hand, Prop.~\ref{prop:densezerocoverage} shows that if $\alpha_0\leq 2$, the coverage probability decays to zero as the network densifies.
This may lead to the pessimistic conjecture that the potential throughput would decrease with the network density.
Fortunately, this is not necessarily true.
A complete characterization of the potential throughput scaling is given in the following theorem.


\begin{theorem}[Throughput Scaling under the Dual-slope Model]
Under the dual-slope path loss model, as $\lambda\to\infty$, the potential throughput $\tau_{l_2(\alpha_0,\alpha_1;\cdot)}(\lambda, T)$
\begin{enumerate}
\item
grows linearly with $\lambda$ if $\alpha_0>2 $,
\label{itm:Tscaling1}
\item
scales sublinearly with rate $\lambda^{2-\frac{2}{\alpha_0}}$ if $1<\alpha_0 < 2 $,
\label{itm:Tscaling2}
\item
decays to zero if $\alpha_0< 1$.
\label{itm:Tscaling3}
\end{enumerate}
\label{thm:TScaling}
\end{theorem}

\begin{IEEEproof}
See Appendix~\ref{app:TScaling}.
\end{IEEEproof}

Due to the technical subtlety, Theorem~\ref{thm:TScaling} does not include the cases of $\alpha_0 =1, 2$
(a slightly different proof technique needs to be tailored exclusively for these points).
Yet, by continuity, we conjecture that the potential throughput
scales linearly at $\alpha_0 = 2$ and converges to some finite value at $\alpha_0=1$.

Theorem~\ref{thm:TScaling} provides theoretical justification to the potential of cell densification
despite the slightly pessimistic results given in Prop.~\ref{prop:densezerocoverage}.
Under the dual slope model, even if $\alpha_0 < 2$ and the coverage probability
goes to zero as the network densifies, the cell splitting gain can still scale up
the potential throughput of the network as long as $\alpha_0 > 1$
which practically holds in most of the cases of interest.

\begin{figure}[tbh]
\centering
\psfrag{lambda}[c][c]{$\lambda$}
\psfrag{T}[c][t]{$\tau_{l_2(\alpha_0,\alpha_1;\cdot)}(\lambda, T)$}
\includegraphics[width=\linewidth]{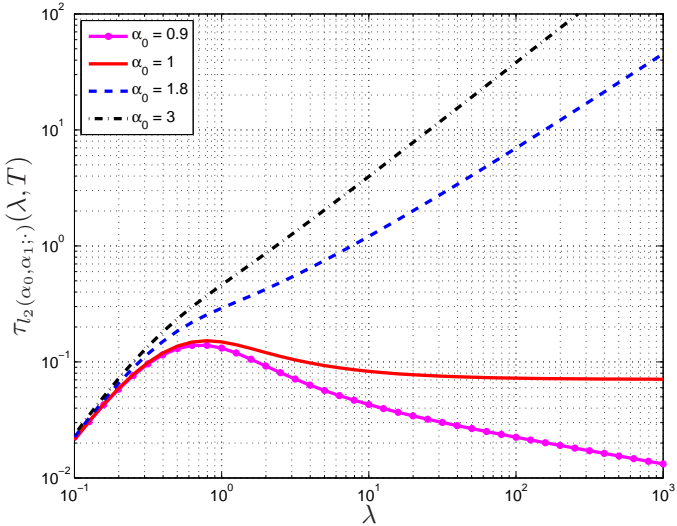}
\caption{Potential throughput scaling with network density.
Here, $\alpha_0 = 0.9,1,1.8,3$, $\alpha_1 = 4$, $R_c=1$, $ \sigma^2 = 0$, $ T = 1$.\label{fig:PTScalingtheta1_R1_lambda0point01_1000_c}}
\end{figure}

Fig.~\ref{fig:PTScalingtheta1_R1_lambda0point01_1000_c} verifies the scaling
results given in Theorem~\ref{thm:TScaling}.
As expected, we observe a \emph{phase transition} at $\alpha_0 = 1$:
if $\alpha_0<1$, the asymptotic potential throughput goes to zero;
if $\alpha_0>1$, it goes to infinity.
For $\alpha_0=1$, numerical results suggest that asymptotic potential throughput
converge to a \emph{positive} finite value.

\section{Multi-slope Path Loss Model\label{sec:Mslope}}

The previous sections have focused on the dual-slope path loss function.
Since Lemma~\ref{lem:CovPGenPL} applies for arbitrary path loss functions (whenever the integral exists),
explicit (integral) expression for the coverage probability of the multi-slope path loss function (Def.~\ref{def:Multi-slopePLM})
can be derived analogous to Theorem~\ref{thm:2exp}.

\begin{theorem}
The coverage probability under the $N$-slope path loss model ($N\geq 3$) is
in \eqref{equ:PcMslp},
where $I_i(\{\alpha_l\},\{R_l\}, T; x)$ is given in \eqref{equ:IiMslp} (both equations are at the top of the next page),
$\delta_i = 2/\alpha_i,~i \in [N-1]\cup\{0\}$ and $C_\beta(x) = {_2F}_1(1,\beta;1+\beta;-x)$.
\label{thm:Nexp}
\end{theorem}
%

\begin{figure*}[!t]
\normalsize
\begin{equation}
\Pc^\SINR_{l_N}(\lambda, T) = 
				\lambda\pi \left( \sum_{i=0}^{N-2}\int_{R_i^2}^{R_{i+1}^2} e^{-\lambda\pi I_i(\{\alpha_l\},\{R_l\}, T; x) 
					 } 
				e^{- T  \sigma^2 x^{\frac{\alpha_i}{2}}/K_i } \d x
+ \int_{R_{N-1}^2}^{\infty} e^{ -\lambda\pi x  C_{-\delta_{N-1}}\left(  T  \right) } e^{- T  \sigma^2 x^{\frac{\alpha_{N-1}}{2}}/K_{N-1}} \d x \right)
\label{equ:PcMslp}
\end{equation}
\begin{multline}
		I_i(\{\alpha_l\},\{R_l\}, T; x) =
						x \left(1-C_{\delta_i}\Big(\frac{1}{ T}\Big)\right)
						+ R_{i+1}^2 C_{\delta_i}\left( \frac{R_{i+1}^{\alpha_i}}{ T x^{\frac{\alpha_i}{2}}} \right)  \\
						+ \sum_{j=i+1}^{N-2} \left(
									R_{j+1}^2 C_{\delta_j}\bigg(\frac{K_i}{K_j} \frac{R_{j+1}^{\alpha_j}}{T x^{\frac{\alpha_i}{2}}}\bigg)
									-R_{j}^2 C_{\delta_j}\bigg(\frac{K_i}{K_j} \frac{R_{j}^{\alpha_j}}{T x^{\frac{\alpha_i}{2}}}\bigg)
									\right) 
						+ R_{N-1}^2 C_{-\delta_{N-1}}\left( \frac{K_{N-1}}{K_i}\frac{T x^{\frac{\alpha_i}{2}}}{R_{N-1}^{\alpha_{N-1}}} \right)
						- R_{N-1}^2
\label{equ:IiMslp}
\end{multline}
\hrulefill
\vspace*{4pt}
\end{figure*}

While Def.~\ref{def:Multi-slopePLM} requires the path loss exponents for the multi-slope path loss function to be increasing,
the proof of Theorem~\ref{thm:Nexp} does not depend on the ordering. 
Thus, Theorem~\ref{thm:Nexp} is true even when $\{\alpha_l\}$ are arbitrarily ordered (so is Theorem~\ref{thm:2exp}).
In the practically important case of ordered path loss exponents (Def.~\ref{def:Multi-slopePLM}), all conclusions drawn in Sect.~\ref{sec:SIR} extend to the multi-slope case. 
In the following theorem, we summarize these main conclusions.

\begin{theorem}
The coverage probability with the multi-slope path loss function given by Def.~\ref{def:Multi-slopePLM}  satisfies the following properties:
\begin{itemize}
\item
$\Pc^\SIR_{l_1(\alpha_0;\cdot)} (\cdot,  T) \lesssim \lim_{\lambda\to\infty} \Pc^\SIR_{l_N} (\lambda,  T) $ (as $\lambda \to \infty$).
\item
$\lim_{\lambda\to 0} \Pc^\SIR_{l_N}(\lambda,  T) \lesssim \Pc^\SIR_{l_1(\alpha_{N-1};\cdot)} (\cdot,  T)$  (as $\lambda \to 0$).
\item
$ \Pc^\SIR_{l_N} (\lambda_1, T) \geq \Pc^\SIR_{l_N}(\lambda_2, T)$, for all $\lambda_1\leq\lambda_2$ and $ T \geq 0$
\end{itemize}
where $\lesssim$ denotes less than equal to and asymptotically equal to, $l_i(\|x\|) = K_i \|x\|^{-\alpha_i}$, for $i\in\{0,1,2,\cdots,N\}$.
\end{theorem}

\begin{figure}[tbh]
\centering
\psfrag{theta}[c][b]{$ T$}
\psfrag{Pc}[c][t]{$\Pc^\SINR_{l_3}$}
\begin{overpic}[width=\linewidth]{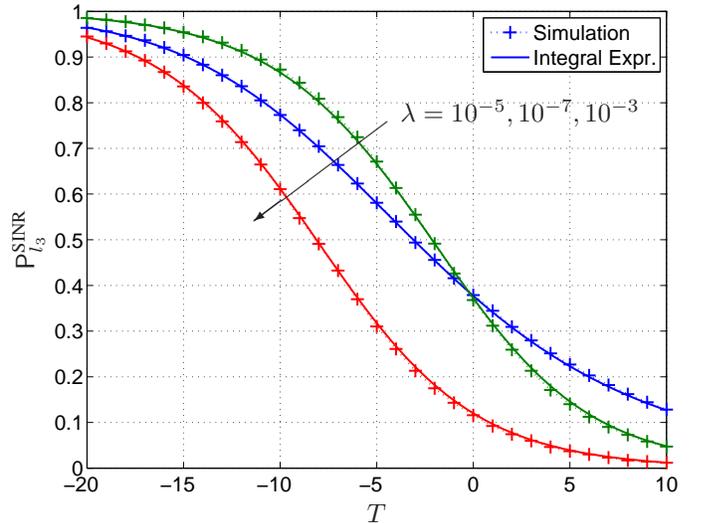}
	\put(55,60){\vector(-4,-3){20}}
	\put(57,60){$ \lambda=10^{-5}, 10^{-7}, 10^{-3}$}
\end{overpic}
\caption{SINR ccdf from simulation and Theorem~\ref{thm:Nexp}.
Here, the number of slopes $N=3$, $[\alpha_0\; \alpha_1\; \alpha_2] = [0\; 2\; 4]$
and $[R_1\; R_2] = [1\; 267]$, $ \sigma^2 = 10^{-8}$.\label{fig:SINRccdfalpha_v_024_Rb_v_1267_lambda1e-07_0.001_W_1e-08_c}}
\end{figure}

In Fig.~\ref{fig:SINRccdfalpha_v_024_Rb_v_1267_lambda1e-07_0.001_W_1e-08_c}, we validate Theorem~\ref{thm:Nexp} with simulations.
We combine the classic two-ray model with a bounded path loss model to create a 3-slope path loss model with $[\alpha_0\; \alpha_1\; \alpha_2] = [0\; 2\; 4]$
and $[R_1\; R_2] = [1\; 267]$.\footnote{Here, we use standard units and $R_2=267$ m comes from the two-ray example mentioned in Sect.~\ref{sec:Intro}.} 
The noise variance is set to $10^{-8}$, corresponding to an $80$ dB \SNR at unit distance. An exact match between analysis and simulation is observed in the figure. Despite the more refined model, similar trends can be observed as in the case with the dual slope model (Fig.~\ref{fig:SINRccdfalpha02_alpha14_lambda0point1_10_c}).

Similarly, following the same proof techniques of those of Theorem~\ref{thm:TScaling},
it is straightforward to generalize the throughput scaling results from the dual-slope
path loss model to the multi-slope path loss model, resulting in the following theorem.

\begin{theorem}[Throughput Scaling under Multi-slope Path Loss Model]
Under the multi-slope path loss model, as $\lambda\to\infty$, the potential throughput $\tau_{l_N}(\lambda, T)$
\begin{enumerate}
\item
grows linearly with $\lambda$ if $\alpha_0>2 $,
\label{itm:MTscaling1}
\item
scales sublinearly with rate $\lambda^{2-\frac{2}{\alpha_0}}$ if $1<\alpha_0 < 2 $,
\label{itm:MTscaling2}
\item
decays to zero if $\alpha_0< 1$.
\label{itm:MTscaling3}
\end{enumerate}
\label{thm:MTScaling}
\end{theorem}

Theorem~\ref{thm:MTScaling} shows that there (still) exists
a phase transition for the asymptotic scaling of network throughput under the multi-slope
path loss model,
and the phase transition happens at the same critical values of $\alpha_0$.
Intuitively, in the ultra-dense regime ($\lambda\to\infty$),
infinitely number of BSs are in the nearest field (subject to path loss exponent $\alpha_0$),
making the scaling independent of $\alpha_n, n\geq 1$.
Nevertheless, the values of $\alpha_n, n\geq 1$ as well as $R_n, n\in [N-1]$
affect the $\SINR$ distribution in the non-asymptotic regime.

\section{Conclusions\label{sec:conclu}}

This paper analyzes cellular network coverage probability and potential throughput under the dual-slope path loss model. We show that despite being a seemingly minor generalization, the dual-slope path loss model produces many surprising observations that stand in sharp contrast to results derived under standard path loss models. In particular, we show the monotonic decrease of $\SIR$ with infrastructure density and the existence of a coverage-maximizing density. Both results are consistent with recent findings based on
 other non-standard path loss functions \cite{RamasamyGantiMadhow2013,BaiHeath2015,BaiVazeHeath2014,BaccelliZhang2015}.

By studying the potential throughput, we show that there exists a phase transition on the asymptotic potential throughput of the network. If the near-field path loss exponent $\alpha_0$ is less than one, the potential throughput goes to zero as the network densifies.
If $\alpha_0>1$, the potential throughput grows (unboundedly) with denser network deployment,
but the growth rate may be sublinear depending on the path loss exponent. Since in most practical cases, we have $\alpha_0>1$, this implies network scalability even without intelligent scheduling.

We believe this paper should lead to further scrutiny of the idealized standard path loss model. The dual-slope and multi-slope path loss functions are important potential substitutes with much more precision and seemingly adequate tractability.
As the cellular network densifies and new technologies are introduced, existing knowledge need to be refined in view of these models. For example, (i) local cell coordination and coordinated multipoint processing (CoMP) may be much more powerful than previously predicted since near-field interferers can produce much stronger interference than far-field ones; (ii) successive interference cancellation (SIC) may be less useful or more dependent on power control since near-by transmitters may produce less differentiable received powers; (iii) in HetNets, closed subscriber groups may be more harmful to nearby users, and the benefit of load balancing may be less than expected due to higher received power from nearby small cells and lower received power from far-away macrocells; and (iv) device-to-device (D2D) communication may be (even) more power-efficient than foreseen due to smaller near-field path loss, but demanding more careful scheduling to mitigate near-field interference.

\section{Acknowledgments}

The authors wish to thank Anthony Soong (Huawei) for the suggestion to investigate dual-slope path loss models and for providing some empirical data supporting their accuracy.  The authors also wish to thank Sarabjot Singh (Nokia) for comments on early drafts of the paper
and sharing his insights on recent millimeter wave research.

\appendices

\section{Proof of Lemma~\ref{lem:SIRbounds}\label{app:SIRbounds}}

\begin{IEEEproof}
Since the lemma states for arbitrary realization $\omega$,
the statistics of the marked point process is not relevant.
Instead of carrying $\omega$ for the rest of the proof,
we will use $h_x$, $x\in\Phi$ to denote $h_x(\omega)$, $x(\omega)\in\Phi(\omega)$ for simplicity.

First, we focus on the first part (first bullet) of the lemma and assume $\SIR_{l_1(\alpha_0;\cdot)}(\omega) >  T$.
The proof proceeds in two cases separately: $\|x^*\|\leq R_c$ and $\|x^*\|>R_c$.
For $\|x^*\|\leq R_c$, we have $l_2(\alpha_0,\alpha_1;x^*) = l_1(\alpha_0;x^*)$.
Since $l_2(\alpha_0,\alpha_1;x)\leq l_1(\alpha_0;x),\; \forall x\neq o$,
we obtain $h_{x^*} l_2(\alpha_0,\alpha_1;x^*) = h_{x^*} l_1(\alpha_0;x^*) >  T \sum_{y\in\Phi\setminus\{x^*\}} h_y l_1(\alpha_0;y) \geq  T \sum_{y\in\Phi\setminus\{x^*\}} h_y l_2(\alpha_0,\alpha_1;y)$,
\ie $\SIR_{l_1(\alpha_0;\cdot)} (\omega) >  T$ implies $\SIR_{l_2(\alpha_0,\alpha_1;\cdot)} (\omega) >  T$.
For $\|x^*\|> R_c$, given $\SIR_{l_1(\alpha_0;\cdot)} (\omega) >  T$, we have
\begin{align*}
h_{x^*} l_2(\alpha_0,\alpha_1;x^*) &= h_{x^*} \eta \|x^*\|^{-\alpha_1} \\
				&= h_{x^*} \eta\|x^*\|^{-\alpha_0} \frac{\|x^*\|^{\alpha_0}}{\|x^*\|^{\alpha_1}} \\
				&\stackrel{\textnormal{(a)}}{>}  T \eta \frac{\|x^*\|^{\alpha_0}}{\|x^*\|^{\alpha_1}} \sum^{\|y\|>\|x^*\|}_{y\in\Phi}  h_y l_1(\alpha_0;y)		\\\displaybreak[0]
				&=  T \sum^{\|y\|>\|x^*\|}_{y\in\Phi} h_y \eta\|y\|^{-\alpha_1} \left(\frac{\|y\|}{\|x^*\|}\right)^{\vartriangle\alpha}	\\
				&\stackrel{\textnormal{(b)}}{>}  T \sum_{y\in\Phi\setminus\{x^*\}} h_y \eta\|y\|^{-\alpha_1} 		\\
				&=  T \sum_{y\in\Phi\setminus\{x^*\}} h_y l_2(\alpha_0,\alpha_1;y),
\end{align*}
where $\eta = R_c^{\vartriangle\alpha}$, $\vartriangle\alpha = \alpha_1 - \alpha_0$, (a) is due to $\SIR_{l_1(\alpha_0;\cdot)} (\omega) >  T$ and (b) comes from the fact that $\|x^*\|<\|y\|,\;\forall y\in\Phi\setminus\{x^*\}$.

The same idea applies to the proof of the second part of the lemma.
To make the proof more strightforward, we first prove that $\SIR_{l_2(\alpha_0,\alpha_1;\cdot)} (\omega) >  T$ implies $\SIR_{\eta l_1(\alpha_1;\cdot)} (\omega) \geq  T$ as follows:
If $\|x^*\|> R_c$, $l_2(\alpha_0,\alpha_1;x) = \eta l_1(\alpha_1;x)$ for all $x_i\in\Phi\cap(x^*,\infty)$
and thus $\SIR_{l_2(\alpha_0,\alpha_1;\cdot)} (\omega) >  T \Longleftrightarrow \SIR_{\eta l_1(\alpha_2;\cdot)} (\omega) \geq  T$.
If $\|x^*\| \leq R_c$, given, $\SIR_{l_2(\alpha_0,\alpha_1;\cdot)} (\omega) >  T$, we have
\begin{align*}
&h_{x^*} \eta l_1(\alpha_1;x^*)=h_{x^*} \eta \|x^*\|^{-\alpha_1}		\\
				&\stackrel{\textnormal{(c)}}{>}  T \eta \|x^*\|^{\vartriangle\alpha} 
								\left(\sum^{\|y\|\leq R_c}_{y\in\Phi\setminus\{x^*\}} h_y \|y\|^{-\alpha_0}
								+ \sum^{\|y\|>R_c}_{y\in\Phi} h_y \eta\|y\|^{-\alpha_1} \right)		\\
				&=  T \left(\sum^{\|y\|\leq R_c}_{y\in\Phi\setminus\{x^*\}} h_y \eta \|y\|^{-\alpha_1} \left(\frac{\|y\|}{\|x^*\|}\right)^{\vartriangle\alpha}				\right. \\
				&\phantom{=}~		+\left.		 \sum^{\|y\|>R_c}_{y\in\Phi} h_y \eta\|y\|^{-\alpha_1} \left(\frac{R_c}{\|x^*\|}\right)^{\vartriangle\alpha}\right)		\\
				&\stackrel{\textnormal{(d)}}{>}  T \sum_{y\in\Phi\setminus\{x^*\}} h_y \eta \|y\|^{-\alpha_1} 		
				=  T \sum_{y\in\Phi\setminus\{x^*\}} h_y \eta l_1(\alpha_1;y),
\end{align*}
where (c) is due to the assumption $\SIR_{l_2(\alpha_0,\alpha_1;\cdot)} (\omega) >  T$ and (d) takes into account the fact that
$\|y\|>\|x^*\|,\;\forall y\in\Phi\setminus\{x^*\}$ and $\|x^*\|\leq R_c$.

Realizing that $\SIR_{ l_1(\alpha_1;\cdot)} (\omega) = \SIR_{k l_1(\alpha_1;\cdot)} (\omega)$ for all $k>0$,
we complete the proof for the second part of the lemma.
\end{IEEEproof}

\section{Proof of Prop.~\ref{prop:densezerocoverage} \label{app:densezerocoverage}}

\begin{IEEEproof}
The following proof is to show $\Pc^\SIR_{l_2}\to 0$ as $\lambda\to\infty$.
Since $\Pc^\SINR_{l_2}\leq\Pc^\SIR_{l_2}$ the same result for SINR coverage follows naturally.

We will first focus on the case of $\alpha_0<2$.
The result of $\alpha=2$ then follows from the continuity of the coverage probability expression \eqref{equ:2expCP}.
Using Lemma ~\ref{lem:CovPGenPL} and setting $W=0$, we can upper bound the coverage probability as follows,
$\Pc^\SIR_{l_2} $
\begin{align*}
						&\pleq{a} \lambda\pi \int_0^\infty \exp\left({{-\lambda\pi y \Bigg(1+\int_1^{1 \vee \frac{R_c^2}{y}}
									\frac{ T}{ T+\frac{l_2(\sqrt{y})}{l_2(\sqrt{t y})}}\d t\Bigg)}} \right)  \d y		\\
						&= \lambda\pi \int_0^{R_c^2} \exp\left({-\lambda\pi y \Big(1+\int_1^{\frac{R_c^2}{y}} 		
									\frac{ T}{ T+t^{\frac{\alpha_0}{2}}}\d t  \Big)}\right)  \d y			\\
						&\phantom{=}~			+ \lambda\pi \int_{R_c^2}^\infty e^{-\lambda\pi R_c^2} \d y		\\
						&\peq{b} {A(\lambda, R_c, \alpha_0,  T)}
									+ e^{-\lambda\pi R_c^2},\numberthis\label{equ:alpha0<2CPUB}
\end{align*}
where $a\vee b = \max\{a,b\}$, ${A(\lambda, R_c, \alpha_0,  T)} \triangleq $
\begin{equation*}	
					{\lambda\pi R_c^2 \int_0^{1} \exp\left({-\lambda\pi u R_c^2  \Big(1+\int_1^{\frac{1}{u}} 
									\frac{ T}{ T+t^{\frac{\alpha_0}{2}}}\d t  \Big)}\right) \d u},
\end{equation*}
(b) is based on the change of variable $y\to u R_c^2$, and (a) is based on
truncating the interval of integration in the exponent with the intuition of ignoring the interference
coming from BSs farther than $R_c$.

Since the second term of \eqref{equ:alpha0<2CPUB} converges to zero with $\lambda\to\infty$,
to prove the lemma we only need to show that $A(\lambda,R_c,\alpha_0, T)$
goes to zero.
For an increasing sequence of $\lambda_n$, let 
\begin{equation}
f_n(x) = \lambda_n \pi R_c^2 \exp\left({-\lambda_n\pi R_c^2 x \left(1+\int_1^{\frac{1}{x}} 
									\frac{ T}{ T+t^{\frac{\alpha_0}{2}}}\d t\right)}\right).
\label{equ:fn}
\end{equation}
It is clear that $f_n(x) \to 0$ almost everywhere on $(0,1)$.
Also, \[0\leq f_n (x) \leq g(x) \triangleq \frac{1}{x e \left(1+\int_1^{\frac{1}{x}} 
									\frac{ T}{ T+t^{\frac{\alpha_0}{2}}}\d t\right)}\]
and it is straightforward to check that $g(x)$ is integrable on $(0,1)$ for $0 \leq \alpha_0<2$.
By the dominated convergence theorem, we have $\lim_{\lambda\to\infty} A(\lambda, R_c, \alpha_0,  T) = 0$ and thus complete the proof.
\end{IEEEproof}

\section{Proof of Lemma~\ref{lem:SIRmono}\label{app:SIRmono}}

\begin{IEEEproof}
Consider a linear mapping $f:\R^2\to\R^2$ such that $f(x) = ax$ for $a>1$.
By the mapping theorem\cite{net:mh12}, it is easy to show that for any homogeneous PPP $\Phi\subset\R^2$
with intensity $\lambda$, $f(\Phi)$ is also a homogeneous PPP on $\R^2$ with intensity $\lambda/a$.
With slight abuse of notation, we let the same mapping operate on the space of the marked PPP (but take effect only on the ground process), \ie $f(\hat\Phi) = f(\{(x_i, h_{x_i})\}) = \{(a x_i, h_{x_i})\}$.
By the same argument, $f(\hat\Phi)$ is a marked PPP with intensity $\lambda/a$
and marked by the same iid fading marks.

If we define the indicator function $\chi_{ T,l} : \R^2\times\R^+ \to \{0,1\}$ as follows:
\begin{equation*}
	\chi_{ T,l}(\hat\Phi) = \left\{
			\begin{array}{ll}
			1,& \textnormal{if  }	h_{x^*}l(x^*)> T \sum_{x\in\Phi\setminus\{x^*\}} h_xl(x)		
			\\
			0,&	\textnormal{otherwise,}
			\end{array}	
	\right.
\end{equation*}
we have $\Pc_l^\SIR(\lambda,  T) = \E \left[ \chi_{ T,l}\left(\hat\Phi(\lambda)\right)\right]$,
where we use $\hat\Phi(\lambda)$ to emphasize that the density of the ground process is $\lambda$.

The key of the proof then comes from the observation that 
$\chi_{ T,l_2}(\hat\phi) \leq \chi_{ T,l_2}\left(f(\hat\phi)\right)$
for all marked point pattern $\hat\phi = \{(x_i, h_{x_i})\}\subset\R^2\times\R^+$ and $a>1$.
More specifically, if $\|x^*\|>R_c$, then $\chi_{ T,l_2}(\hat\phi) = 1$ implies 
$	h_{x^*} \|x^*\|^{-\alpha_1} >  T\sum_{x\in\phi\setminus\{x^*\}} h_x \|x\|^{-\alpha_1}$.
Multiplying both sides of the inequality by $a^{-\alpha_1},\; a>1$ leads to the conclusion
that $\chi_{ T,l}\left(f(\hat\phi)\right) = 1$.
If $\|x^*\|<R_c$, we need to separate two cases: $a\|x^*\| \leq R_c$ and $a\|x^*\| > R_c$.
In the former case, $l_2(a x^*) = a^{-\alpha_0} l_2(x^*)$, thus $\chi_{ T,l_2}(\hat\phi) = 1$ implies
\begin{multline*}
h_{x^*}l_2(ax^*) = h_{x^*} a^{-\alpha_0}l_2(x^*) 			\\
								>  T\left(\sum_{x\in\phi\setminus\{x^*\}}^{\|x\|\leq\frac{R_c}{a}} h_x a^{-\alpha_0} l_2(x)	 
								+ \sum_{x\in\phi}^{\|x\|\in(\frac{R_c}{a},R_c]} h_x a^{-\alpha_0} l_2(x)				\right.	\\ \left.
								+ \sum_{x\in\phi}^{\|x\|>R_c} h_x a^{-\alpha_0} l_2(x)\right),
\end{multline*}
where for $\|x\|\in(\|x^*\|,\frac{R_c}{a}]$, we have $a^{-\alpha_0}l_2(x) = l_2(x)$; 
for $\|x\|\in(\frac{R_c}{a},R_c]$,
we have $a^{-\alpha_0}l_2(x) = \left(\frac{x}{R_c/a}\right)^{\vartriangle\alpha} l_2(ax) \geq l_2(ax)$,
where $\vartriangle\alpha = \alpha_1-\alpha_0$;
for $\|x\|>R_c$, we have $a^{-\alpha_0}l_2(x) = a^{-\alpha_0} \eta x^{-\alpha_1}  \geq \eta (ax)^{-\alpha_1} = l_2(ax)$
since $a>1$ and $\vartriangle\alpha>0$.
These observations lead to the conclusion that $h_{x^*} l_2(ax^*)>  T\sum_{x\in\phi\setminus\{x^*\}} h_x l_2(ax)$,
\ie $\chi_{ T,l_2}\left(f(\hat\phi)\right) = 1$.
In the latter case where $a\|x^*\|>R_c$, if $\chi_{ T,l_2}(\hat\phi) = 1$, we have
\begin{align*}
&h_{x^*}l_2(ax^*) = h_{x^*} \eta (ax^*)^{-\alpha_1} = h_{x^*} l_2(x) \frac{\eta a^{-\alpha_1}}{\|x^*\|^{\vartriangle\alpha}}	\\
							&\stackrel{\textnormal{(a)}}{>}  T \frac{\eta a^{-\alpha_1}}{\|x^*\|^{\vartriangle\alpha}} 
												\left(\sum_{x\in\phi\setminus\{x^*\}}^{\|x\|\leq R_c} h_x l_2(x) + \sum_{x\in\phi}^{\|x\|> R_c} h_x l_2(x) \right)	\\
												&=  T \left(\sum_{x\in\phi\setminus\{x^*\}}^{\|x\|\leq R_c}  h_x \eta \|a x\|^{-\alpha_1} 
													\left(\frac{\|x\|}{\|x^*\|}\right)^{\vartriangle\alpha} 		\right. \\ 
												&\phantom{=}~\left.	+ \sum_{x\in\phi}^{\|x\|> R_c} h_x \eta \|a x\|^{-\alpha_1} 
													\left(\frac{R_c}{\|x^*\|}\right)^{\vartriangle\alpha} \right)		\\
												&\stackrel{\textnormal{(b)}}{>} T 
												\left(\sum_{x\in\phi\setminus\{x^*\}}^{\|x\|\leq R_c} h_x \eta \|a x\|^{-\alpha_1} 
													+ \sum_{x\in\phi}^{\|x\|> R_c} h_x \eta \|a x\|^{-\alpha_1} 
													 \right)			\\
												&= T \sum_{x\in\phi\setminus\{x^*\}} h_x l_2(ax) \numberthis\label{equ:lax>laxC3}
\end{align*}
where (a) is the due to the assumption $\chi_{ T,l_2}(\hat\phi) = 1$,
and (b) takes into account the assumption $ax^*>R_c$ and $a>1$.
\eqref{equ:lax>laxC3} again leads to $\chi_{ T,l_2}\left(f(\hat\phi)\right) = 1$.

Therefore, 
\begin{multline*}
\Pc_{l_2}^\SIR(\lambda,  T) = \E \left[\chi_{ T,l_2}\left(\hat\Phi(\lambda)\right)\right] 
\\
\leq \E \left[ \chi_{ T,l_2}\left(f(\hat\Phi(\lambda))\right)\right] \peq{c} \E \left[\chi_{ T,l_2}\left(\hat\Phi(\lambda/a)\right)\right],
\end{multline*}
where (c) comes from the fact that $f(\hat\Phi(\lambda))$ is
a marked homogeneous PPP with intensity $\lambda/a$ and with the same iid mark distribution as that of $\hat\Phi(\lambda)$.
\end{IEEEproof}

\section{Proof of Prop.~\ref{prop:SINRcovLB}\label{app:SINRcovLB}}

\begin{IEEEproof}
We start from Lemma~\ref{lem:CovPGenPL}. The $\SINR$
coverage probability can be written as $\Pc^\SINR_{l_2(2,4;\cdot)}(\lambda,T) =$
\begin{multline}
					\underbrace{\lambda \pi \int_0^{R_c^2} e^{{-\lambda\pi y \left(1+\int_1^\infty 
									\frac{T}{T+{y^{-1}}/{l_2(2,4;\sqrt{t y})}}\d t\right)} - \sigma^2 T y} \d y}_{A} \\
						+ \underbrace{\lambda \pi \int_{R_c^2}^\infty e^{{-\lambda\pi y \left(1+\int_1^\infty 
									\frac{T}{T+ R_c^2 y^{-1}/{l_2(2,4;\sqrt{t y})}}\d t\right)-\sigma^2 T y^2/ R_c^2}} \d y}_{B},
\label{equ:SINRCov}
\end{multline}
where $A$ ($B$, resp.) is the probability that the user being covered by an BS closer (farther, resp.) than $R_c$.
$B$ can be simplified (with the change of variable $y\to x R_c^2$, as in Thm.~\ref{thm:2exp}) into
\begin{multline*}
\lambda\pi R_c^2 \int_1^{\infty} \exp\left( -\lambda\pi R_c^2 x C_{-\frac{1}{2}}(T)\right) e^{-T \sigma^2 x^2 R_c^2} \d x
=	\\
		\sqrt{\pi}\rho_1(\lambda,T,\sigma^2) e^{\frac{1}{4}\left({C_{-\frac{1}{2}}(T)}\rho_1(\lambda,T,\sigma^2)\right)^2} 	\\
					\times \Q\left(\frac{1}{\sqrt{2}}\rho_1(\lambda,T,\sigma^2) C_{-\frac{1}{2}}(T) + \sqrt{2T \sigma^2} R_c\right).
\end{multline*}
Thus, to prove the proposition, it is just to lower bound $A$ of \eqref{equ:SINRCov}.

We first focus on the exponent inside the integral and observe that
\begin{align*}
	&\int_1^\infty \frac{T}{T+{y^{-1}}/{l_2(2,4;\sqrt{t y})}}\d t \\
		&= \int_1^{\frac{R_c^2}{y}} \frac{T}{T+{t}}\d t
			+ \int_{\frac{R_c^2}{y}}^\infty \frac{T}{T+ t^2 y/R_c^2}\d t \\
		&\pleq{a} T \int_1^{\frac{R_c^2}{y}} \frac{1}{t} \d t
			+ T R_c^2 \frac{1}{y}  \int_{\frac{R_c^2}{y}}^\infty \frac{1}{t^2} \d t		\\
		&= T \log\left(\frac{y}{R_c^2}\right) + T.  \numberthis\label{equ:intUB24}
\end{align*}
Applying \eqref{equ:intUB24} and a change of variables $y \to x R_c^2$,
we obtain
\begin{equation}
	A \geq \lambda \pi R_c^2 \int_0^1 e^{\lambda\pi R_c^2 T x\log(x)} e^{ - \rho_1(\lambda,T,\sigma^2) R_c^2 x} \d x,
\end{equation}
which can be viewed as $K_1 \E[e^{\lambda\pi R_c^2 T X\log(X)}]$
for random variable $X$ with pdf $f_X(x) = K_2 e^{ - \rho_1(\lambda,T,\sigma^2) R_c^2 x}$
($K_1,K_2\in\R^+$ are normalization factors).
Since $e^{x}$ is convex, we apply Jensen's inequality 
\begin{equation}
K_1 \E[e^{\lambda\pi R_c^2 T X\log(X)}]\geq K_1 e^{\lambda\pi R_c^2 T \E[X\log(X)]}
\label{equ:Jensenex}
\end{equation}
and obtain the desired bound.

To see the asymptotic tightness as $\lambda\to 0$,
we can examine the alternative representation of \eqref{equ:SINRCov} in \eqref{equ:2expCP}
and make the following observation which essentially generalizes Fact~\ref{fact:lambdaR2} to
the noisy case: letting $\lambda\to 0 $ (but keeping $R_c$ and $\sigma^2$ fixed) produces the same
effect on $P^\SINR_{l_2}$ as letting $R_c\to 0$ but keeping $\sigma^2 R_c^{\alpha_0}$ and $\lambda$ fixed.
Due to the physical meaning of $A$ and $B$, this implies $B$ dominates $A$ in \eqref{equ:SINRCov} as $\lambda\to 0$.
Since $B$ is exact in the lower bound (in \eqref{equ:SINRCov}, we only lower bounded $A$.), the bound is tight as $\lambda\to 0$.

The asymptotic tightness as $T \to 0$ is observed by examining the (only)
two inequality applied in the derivation: (a) in \eqref{equ:intUB24} and the Jensen's inequality in \eqref{equ:Jensenex}.
Both are tight as $T\to 0$.
\end{IEEEproof}

\section{Proof of Theorem~\ref{thm:TScaling}\label{app:TScaling}}

\begin{IEEEproof}
Since as network density goes to infinity the network becomes interference limited,
it suffices to consider only the case where $W=0$ and the result holds even with noise.

\ref{itm:Tscaling1}) comes directly from Lemma~\ref{lem:capacityLinearScaling}.
To show \ref{itm:Tscaling3}), one could use the same techinques in the proof
of Prop.~\ref{prop:densezerocoverage} thanks to the simple relation between
coverage probability and the potential throughput.
Basically multiplying both sides of \eqref{equ:alpha0<2CPUB} by $\lambda$ gives an upper bound
on the coverage density, \ie $\mu_{l_2}(\lambda, T) \leq \lambda A(\lambda, R_c, \alpha_0,  T) + \lambda\exp(-\lambda\pi R_c^2)$,
where the second term goes to zero as $\lambda\to\infty$.
The first term can also be shown to converge to zero by the dominated convergence theorem.
In particular, using similar construction to \eqref{equ:fn}, $\lambda_n f_n(\cdot)$ goes to
zero almost everywhere and is upperbounded by
\[
	g'(x) = 4/\left( \pi R_c^2 x^2 \left(1+\int_1^{\frac{1}{x}} 
									\frac{ T}{ T+t^{\frac{\alpha_0}{2}}}\d t\right)^2\right),
\]
which is integrable on $(0,1)$ if $\alpha_0 <1$.

To prove \ref{itm:Tscaling2}), we focus on showing that $\Pc^\SIR_{l_2}(\lambda, T) = \Omega (\lambda^{1-\frac{2}{\alpha_0}})$
as $\lambda\to\infty$ given $1\leq \alpha_0 <2$.
We start from Lemma~\ref{lem:CovPGenPL}.
By truncating the (outer) infinite integral to only $(0,R_c^2)$,
we have a lower bound on the coverage probability $\Pc^\SIR_{l_2}(\lambda, T) \geq$
\begin{equation}	
					\lambda \pi \int_0^{R_c^2} \exp\left({-\lambda\pi y \Big(1+\int_1^\infty 
									\frac{ T}{ T+{y^{-\frac{\alpha_0}{2}}}/{l_2(\sqrt{t y})}}\d t\Big)}\right) \d y
\label{equ:PcLB}
\end{equation}
which is essentially the probability
that the typical user being covered by a BS within distance $R_c$.
Further,
\begin{align*}
	&\int_1^\infty \frac{ T}{ T+{y^{-\frac{\alpha_0}{2}}}/{l_2(\sqrt{t y})}}\d t \\
		&= \int_1^{\frac{R_c^2}{y}} \frac{ T}{ T+{t^{\frac{\alpha_0}{2}}}}\d t
			+ \int_{\frac{R_c^2}{y}}^\infty \frac{ T}{ T+ t^{\frac{\alpha_1}{2}} y^{\frac{\alpha_1-\alpha_0}{2}}/\eta}\d t \\
		&\leq  T \int_1^{\frac{R_c^2}{y}} t^{-\frac{\alpha_0}{2}}\d t
			+  T \eta y^{-\frac{\alpha_1-\alpha_0}{2}}  \int_{\frac{R_c^2}{y}}^\infty t^{-\frac{\alpha_1}{2}} \d t		\\
		&= -\frac{2 T}{2-\alpha_0} + \frac{2 T R_c^{2-\alpha_0} (\alpha_1-\alpha_0)}{(2-\alpha_0)(\alpha_1-2)} y^{\frac{\alpha_0}{2}-1}.
\end{align*}
This leads to a simplification of the lower bound in \eqref{equ:PcLB}.
After a change of variable $y\to x R_c^2$, we obtain $\Pc^\SIR_{l_2}(\lambda, T) \geq$
\begin{align*}
 \lambda\pi R_c^2 \int_0^1 e^{-\lambda\pi R_c^2 \left(1-\frac{2 T}{2-\alpha_0}\right)x}
			e^{-\lambda\pi R_c^2 \frac{2 T(\alpha_1-\alpha_0)}{(2-\alpha_0)(\alpha_1-2)}x^\frac{\alpha_0}{2}}\d x,
\end{align*}
which can be lower bounded for $ T\in(0,1-\frac{\alpha_0}{2})$ and $ T\in[1-\frac{\alpha_0}{2},\infty)$ separately.
If $ T\in(0,1-\frac{\alpha_0}{2})$, we have $1-\frac{2 T}{2-\alpha_0}>0$ and
\begin{align*}
			\Pc^\SIR_{l_2}(\lambda, T)
			&\pgeq{a} \lambda\pi R_c^2 \int_0^1 e^{-\lambda\pi R_c^2 \left(1+\frac{2 T}{\alpha_1-2}\right) x^\frac{\alpha_0}{2}} \d x		\\
			& = \delta_0 \frac{(\lambda\pi R_c^2)^{1-\delta_0}}{\left(1+\frac{2 T}{\alpha_1-2}\right)^{\delta_0}}
						\gamma\left(\delta_0, \lambda \pi R_c^2 \Big(1+\frac{2 T}{\alpha_1-2}\Big)\right),
\end{align*}
where $\delta_0=2/\alpha_0$, (a) is due to the fact that $x \leq x^{\alpha_0/2}$ for $0<x<1$ and $\alpha_0\leq 2$
and $\gamma(t,z) = \int_0^z x^{t-1} e^{-x} \d x$ is the \emph{lower} incomplete gamma function.
If $ T\in[1-\frac{\alpha_0}{2},\infty)$, we have
\begin{multline*}
\Pc^\SIR_{l_2}(\lambda, T)
				\geq \lambda\pi R_c^2 \int_0^1
					e^{-\lambda\pi R_c^2 \frac{2 T(\alpha_1-\alpha_0)}{(2-\alpha_0)(\alpha_1-2)}x^\frac{\alpha_0}{2}}\d x		\\
			= \delta_0 \frac{(\lambda\pi R_c^2)^{1-\delta_0}}{\left(\frac{2 T(\alpha_1-\alpha_0)}{(2-\alpha_0)(\alpha_1-2)}\right)^{\delta_0}}
						\gamma\left(\delta_0, \lambda \pi R_c^2 \Big(\frac{2 T(\alpha_1-\alpha_0)}{(2-\alpha_0)(\alpha_1-2)}\Big)\right).
\end{multline*}
Since 
\begin{multline}
\lim_{\lambda\to\infty}\gamma\left(\delta_0, \lambda \pi R_c^2\big(1+\frac{2 T}{\alpha_1-2}\big)\right)				\\
= \lim_{\lambda\to\infty} \gamma\left(\delta_0, \lambda \pi R_c^2 \big(\frac{2 T(\alpha_1-\alpha_0)}{(2-\alpha_0)(\alpha_1-2)}\big)\right) = \gamma(\delta_0),
\end{multline}
we have $\Pc^\SIR_{l_2} (\lambda, T) = \Omega(\lambda^{1-\delta_0})$ for all $ T>0$,
and thus $\tau_{l_2}(\lambda, T) = \Omega(\lambda^{2-\delta_0}) = \mu_{l_2}(\lambda, T)$.
\end{IEEEproof}

\end{document}